\documentclass[10pt,twocolumn]{IEEEtran} 

\usepackage{amsmath,amssymb,mathtools}
\usepackage{stmaryrd}
\usepackage{enumerate}
\usepackage{array}
\usepackage{eqparbox}
\usepackage{frame}
\usepackage{float}
\usepackage{algorithm}
\usepackage{color}
\usepackage[tight,footnotesize]{subfigure}
\usepackage{url}
\usepackage{epsfig}
\usepackage{multirow}
\usepackage{color}
\usepackage{subfigure}
\usepackage{cite}
\usepackage{todonotes}
\usepackage[font=small]{caption}
\usepackage{balance}
\usepackage{graphicx}
\usepackage{graphbox}
\usepackage[export]{adjustbox}
\usepackage{booktabs}
\usepackage{algorithm}
\usepackage{algorithmic}
\usepackage{cases}
\usepackage{tikz}
\usepackage{adjustbox}
\usetikzlibrary{graphs}
\tikzstyle{arrow} = [->,>=stealth]

\newcommand{\real}{\ensuremath{\mathbb{R}}}
\newcommand{\s}{\ensuremath{\mathbb{S}}}
\newcommand{\ltwo}{\ensuremath{\mathbb{L}^2}}

\newcommand{\tabincell}[2]{\begin{tabular}{@{}#1@{}}#2\end{tabular}}

\newtheorem{definition}{Definition}

\begin{document}

	\title{Dynamic Shape Modeling to Analyze Modes of Migration During Cell Motility}
	\author{Ximu Deng$^{\dagger}$, Rituparna Sarkar$^{\dagger}$, \IEEEmembership{Member, IEEE }, Elisabeth Labruyere, \\ Jean-Christophe Olivo-Marin, \IEEEmembership{Fellow, IEEE } 
		\& Anuj Srivastava, \IEEEmembership{Fellow, IEEE } 
        \thanks{Corresponding author: R. Sarkar, email: rituparna.sarkar@pasteur.fr}
		\thanks{$^{\dagger}$ X. Deng and R. Sarkar are joint first authors.}
		\thanks{X. Deng and A. Srivastava are affiliated to Department of Statistics, Florida State University, Tallahassee, FL, USA. (email: xd15@my.fsu.edu, anujsri1968@gmail.com)}. 
		\thanks{R. Sarkar,  E. Labruyere and J.-C. Olivo-Marin are affiliated to Bioimage Analysis Unit, Institut Pasteur, Paris, France (e-mail: (rituparna.sarkar, elisabeth.labruyere, jcolivo)@pasteur.fr).}
	}
	\maketitle
	
	\begin{abstract}	
		This paper develops a generative statistical model for representing, modeling, and comparing the morphological evolution of biological cells undergoing motility. It uses the elastic shape analysis to separate cell kinematics (overall location, rotation, speed, etc.) from its morphology and represents morphological changes using transported square-root vector fields (TSRVFs). This TSRVF representation, followed by a PCA-based dimension reduction, provides a convenient mathematical representation of a shape sequence in the form of a Euclidean time series. Fitting a vector auto-regressive (VAR) model to this TSRVF-PCA time series leads to statistical modeling of the overall shape dynamics. We use the parameters of the fitted VAR model to characterize morphological evolution. We validate VAR models through model comparisons, synthesis, and sequence classifications. For classification, we use the VAR parameters in conjunction with different classifiers: SVM, Random Forest, and CNN, and obtain high classification rates. Extensive experiments presented here demonstrate the success of the proposed pipeline. These results are the first of the kind in classifying cell migration videos using shape dynamics. 
	\end{abstract}
	

	
	\section{Introduction}
	
	\IEEEPARstart{U}{nderstanding} cell motility is crucial to understanding such critical biological phenomena as development, reproduction, healing, and cancer metastasis, among others. Cell motility is a complex process with many simultaneous events, including certain intra- and extra-cellular physiochemical events, that lead to a restructuring of the cytoskeleton~\cite{boquet2021bioimage}. This restructuring, in turn, induces morphological and positional changes in cells causing motility  and complex migration patterns. Under cursory visual inspections, the cell migrations may appear chaotic and random. However, scientists hypothesize that biological influences force certain characteristic structures on motility patterns. The cell motility has two components:  (1)  {\it kinematics}, relating to the gross movement (position, speed, rotations, {\it etc.}) of the cell, and (2) {\it morphology},  relating to changes in its shape. In this paper, we focus on the {\it morphology} of cells to examine dominant modes of cell motility under different intra- and extra-cellular conditions.
	

	 The development of precise mathematical representations and statistical modeling of shapes, especially in cellular biology, is an active area of research. Past research has mainly dealt with either static shapes \cite{dufour2014signal, tweedy2013distinct}, or simply the overall kinematics of single cells \cite{campos2010persistent, miyoshi2003characteristics} primarily ignoring their shapes.  The main challenges in modeling shape dynamics are: How to separate shape variables from kinematic variables during cell motility? How to mathematically represent and statistically model the time series of shapes of cells? How to use these models in characterizing and classifying different modes of cell motility?  In this paper, we develop a systematic, detailed approach for analyzing {\it dynamics of shapes} during cell motility. To the best of our knowledge, this paper is amongst the first to develop such a framework.

	From a biological perspective, the focus here is on unicellular amoebas and some cancer cells. These organisms exhibit amoeboid migration patterns during the invasion of their surroundings or escape from the originating tissues. Such amoeboid motions are characterized by alternate protrusions (or {\it blebs}) and retractions of the cell membranes, which, in turn, result from forces generated by intra-cellular pressure changes and adhesion to the extracellular substrate. Consequently, it becomes interesting to focus on membrane deformations as representing shape dynamics during cell migrations. Specifically, we focus on single-cell \textit{Entamoeba histolytica} as the prototype organism to study cell motility. We will consider amoeba as 2D objects represented by their membranes or boundaries (simple, closed, planar curves) in video frames, thus generating a time series of contours. 

    Observed cell shape evolutions are complex and demonstrate unpredictable irregular pattern. 
     A robust approach that can capture dynamics of the shape of a cell membrane is essential for studying different modes of cell migration. A major challenge is to accurately capture the spatiotemporal behavior of the deforming membrane {\it i.e.}, develop a model to represent the whole sequence of evolving shapes.
	Our main objective is to develop a comprehensive pipeline for extracting shapes from contour sequences, transforming shapes into a Euclidean time series, using statistical models for analyzing such time series, and using model parameters to characterize original shape dynamics. Such a pipeline can be used in various applications ranging from dynamic shape prediction, shape sequence synthesis/simulation, clustering and classification.    
	
    \begin{figure*}
	\begin{center}
	\includegraphics[height=0.25\linewidth]{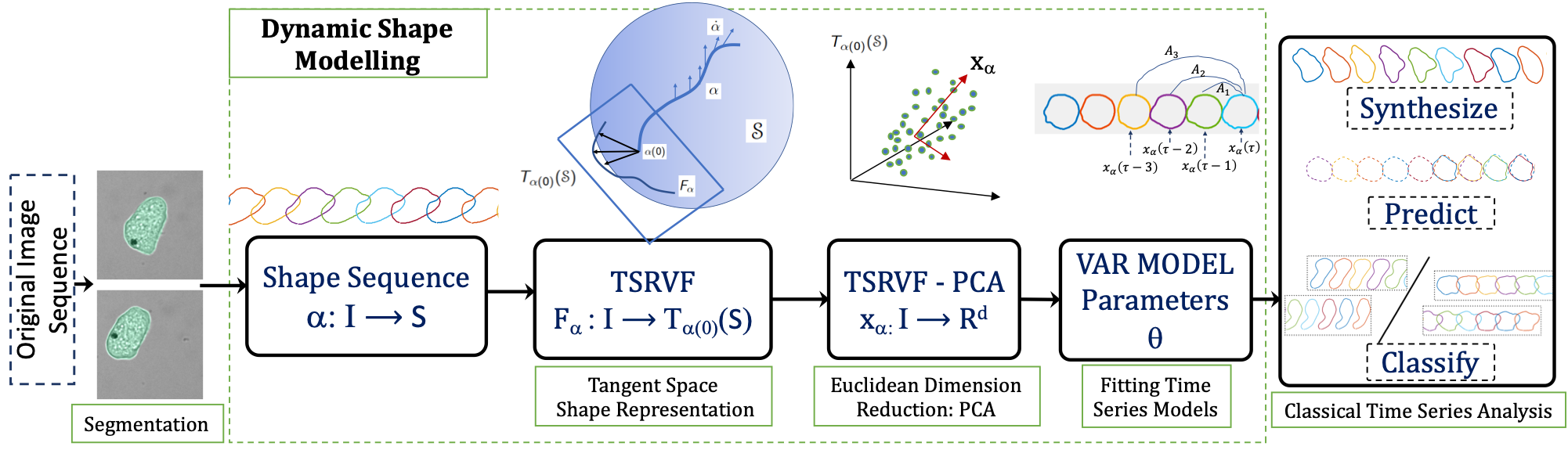}
	    \caption{An overview of the proposed pipeline for characterizing and analyzing dynamics of cellular shapes during motility.} \label{fig:pipeline}
	\end{center}
	\end{figure*}
	\vspace{-2pt}
	
		\begin{figure}[t]
		\begin{center}
			\renewcommand{\tabcolsep}{0.01cm}
			\begin{tabular}{cccc}
			\includegraphics[height=0.20\linewidth]{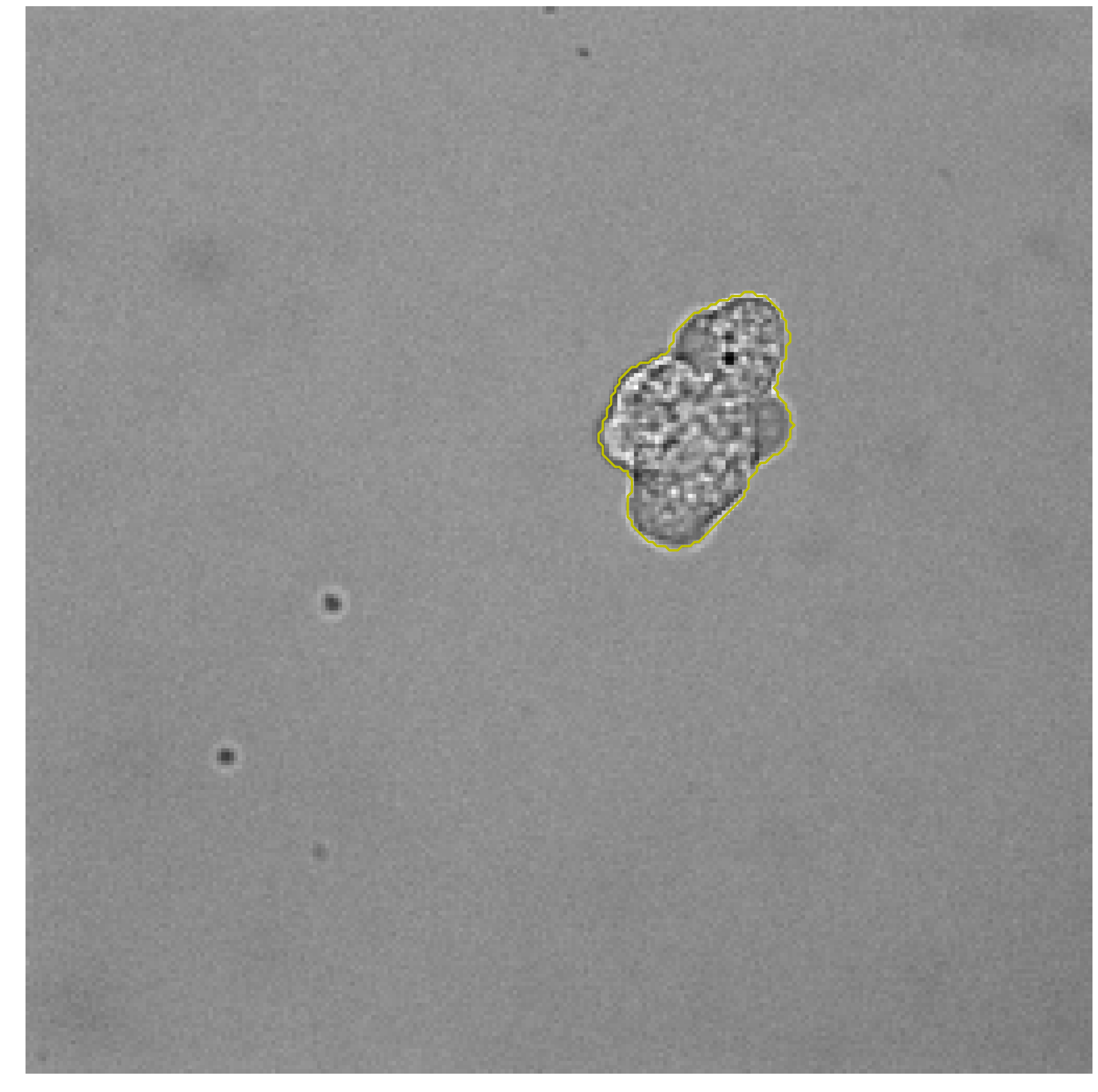}&
			\includegraphics[height=0.20\linewidth]{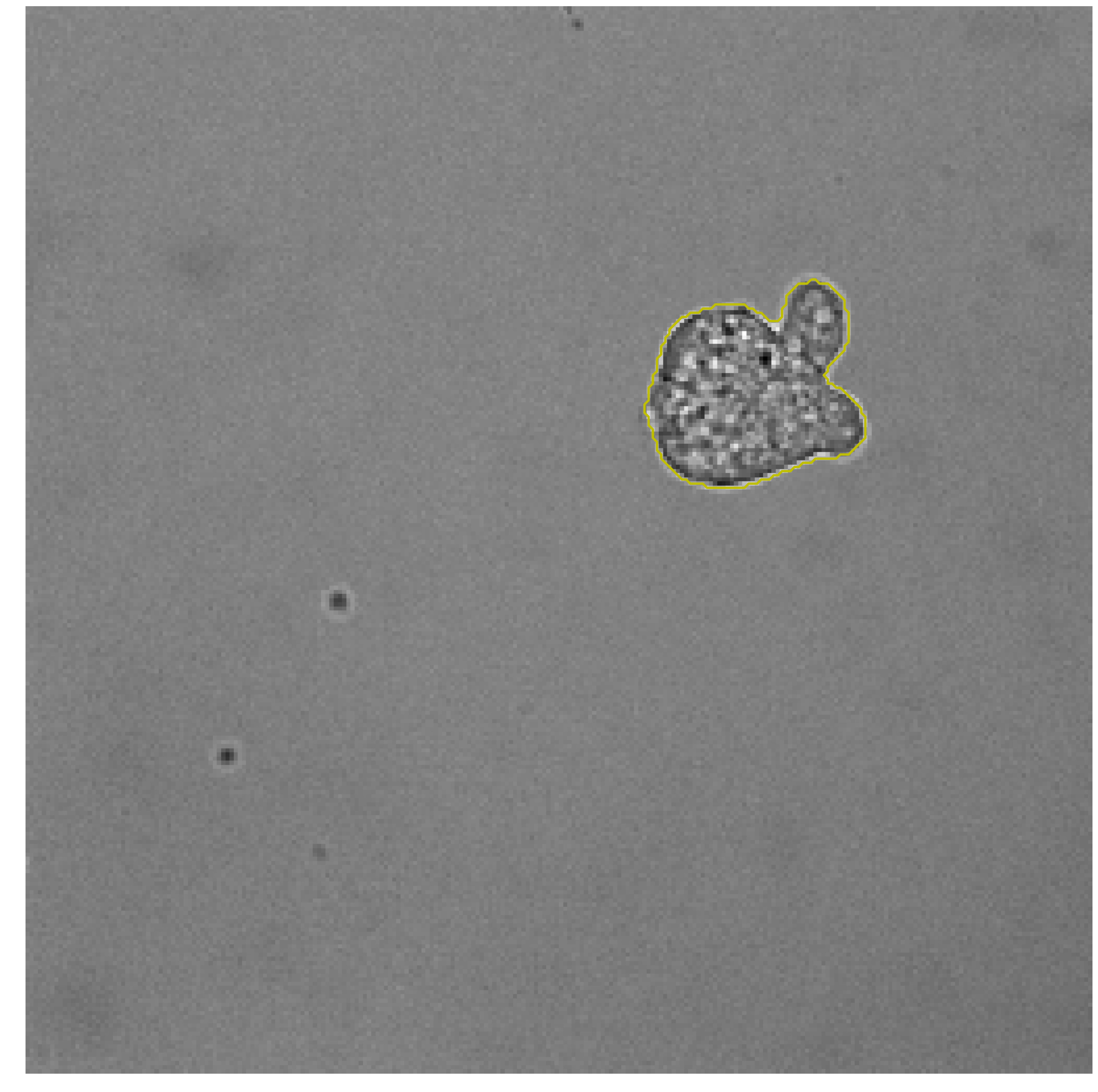} &
			\includegraphics[height=0.2\linewidth]{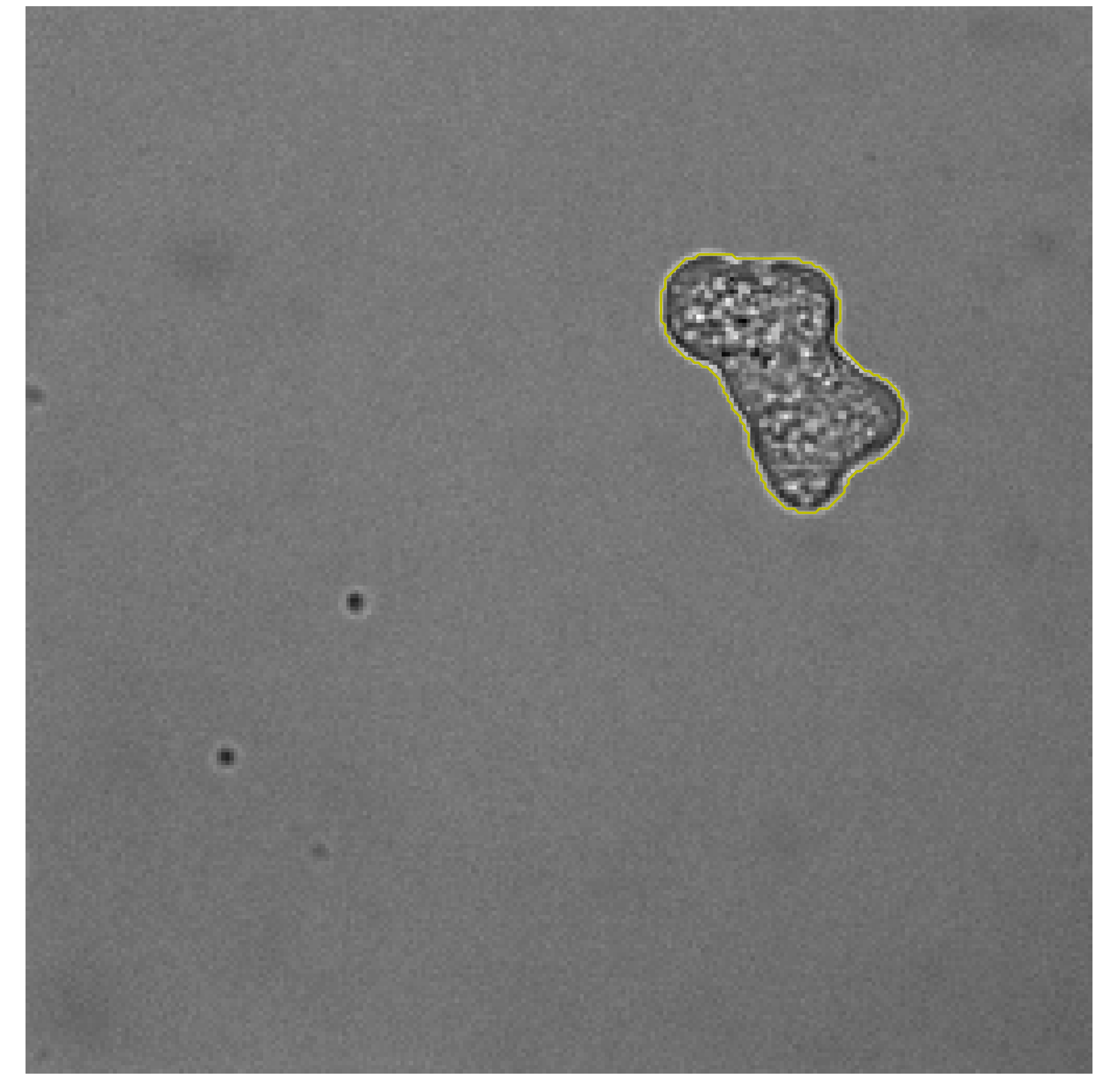} &
			\includegraphics[height=0.20\linewidth]{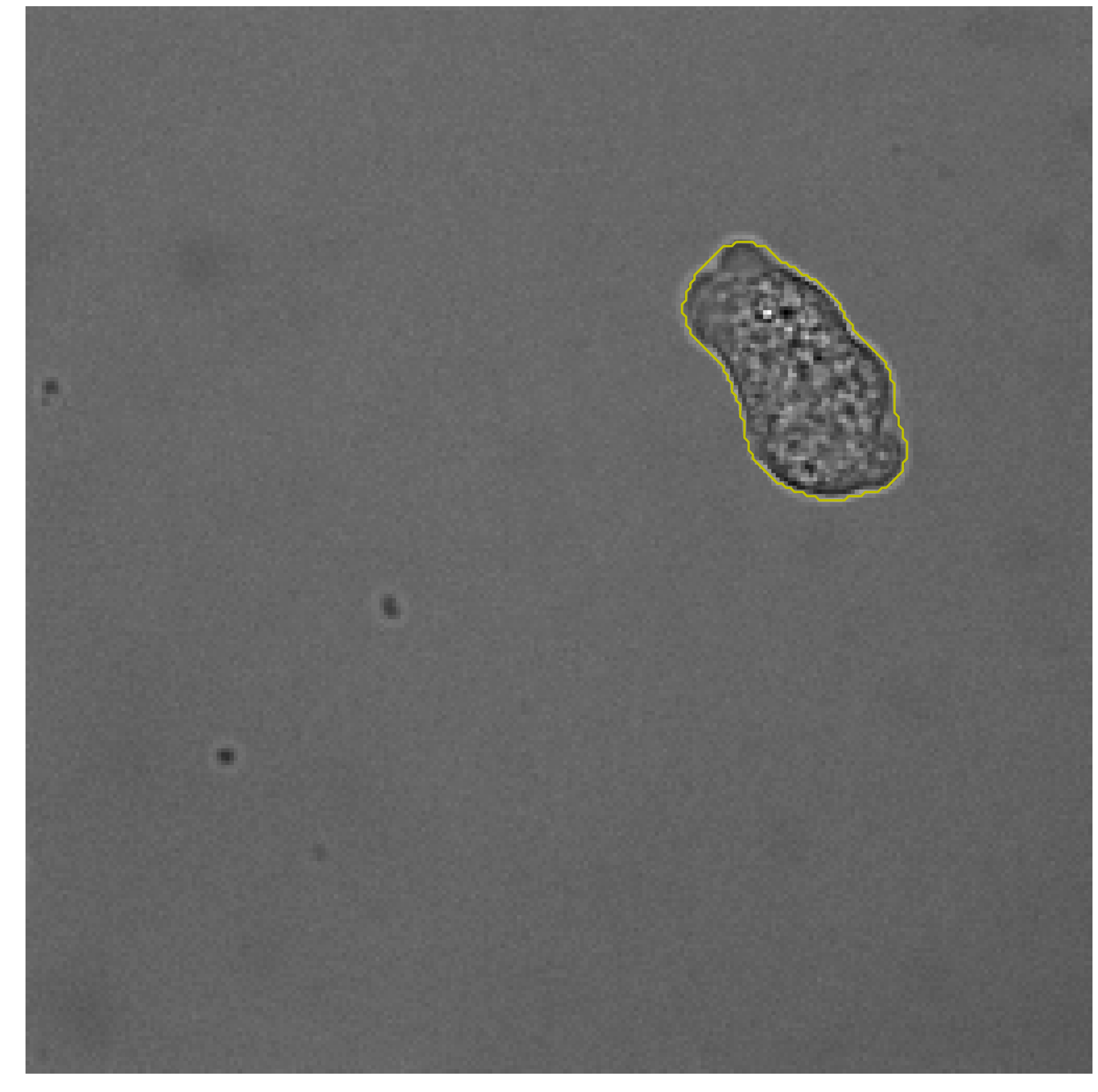} 
			\\
			\includegraphics[height=0.20\linewidth]{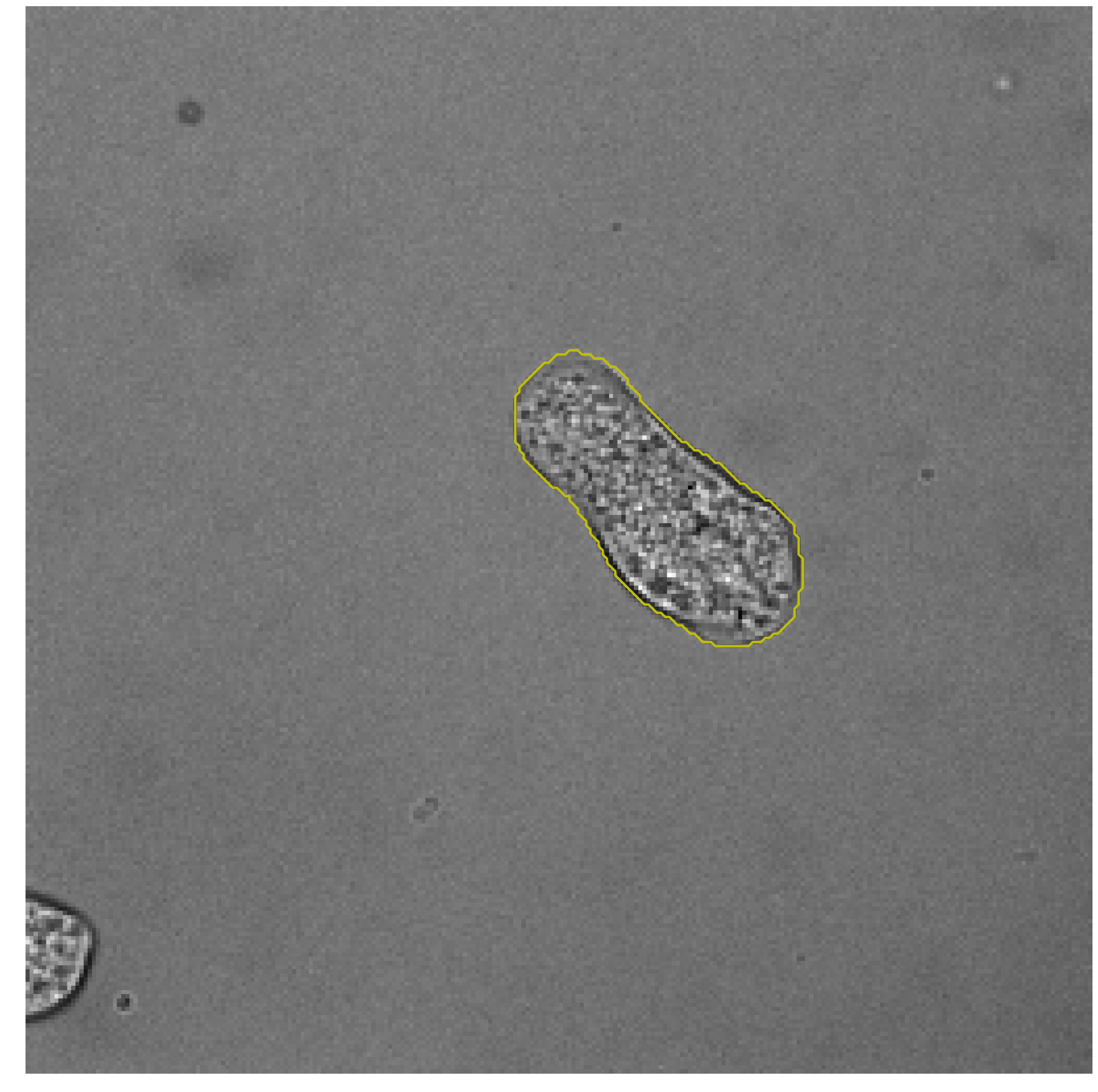}&
			\includegraphics[height=0.20\linewidth]{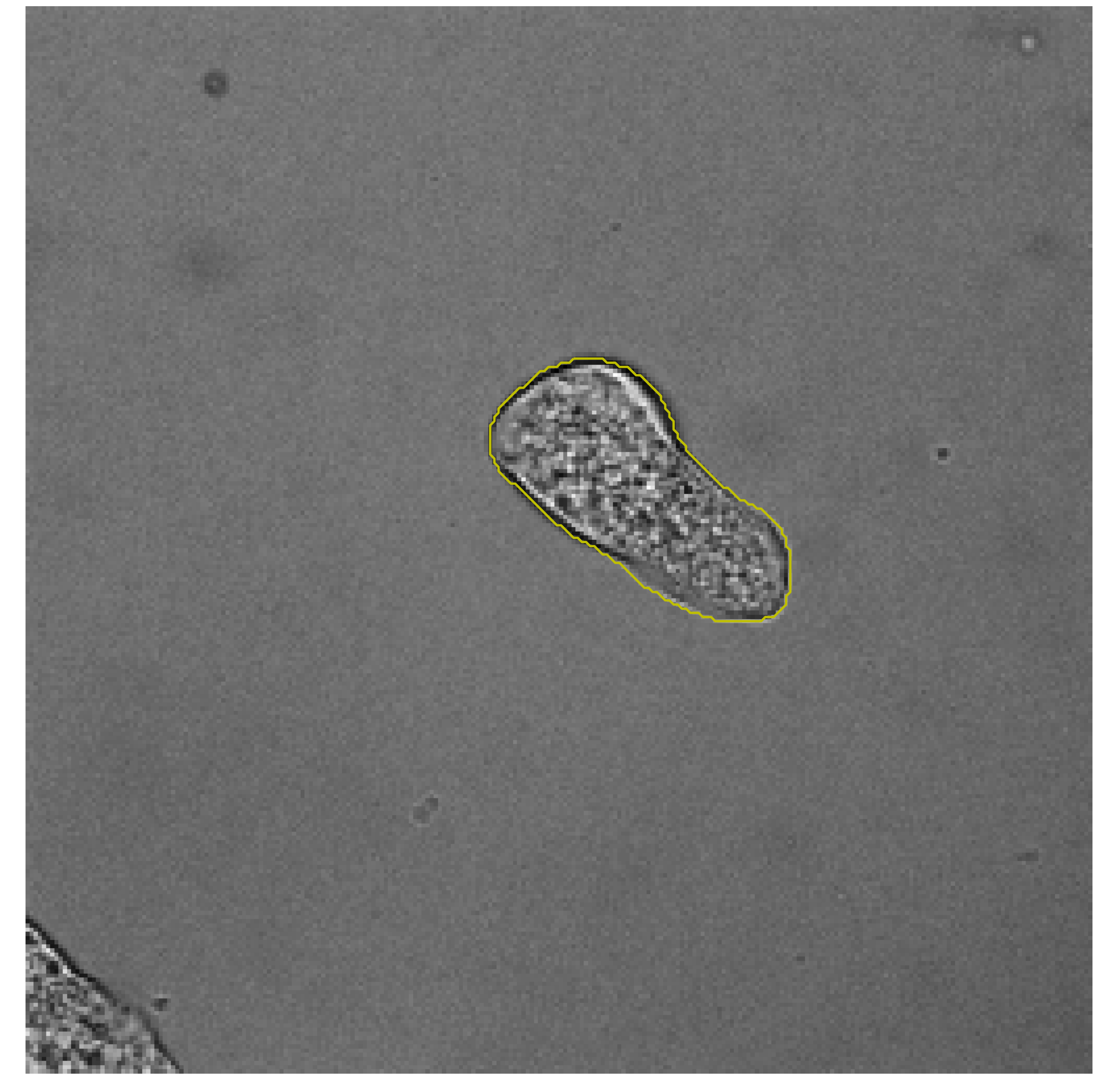}&
			\includegraphics[height=0.20\linewidth]{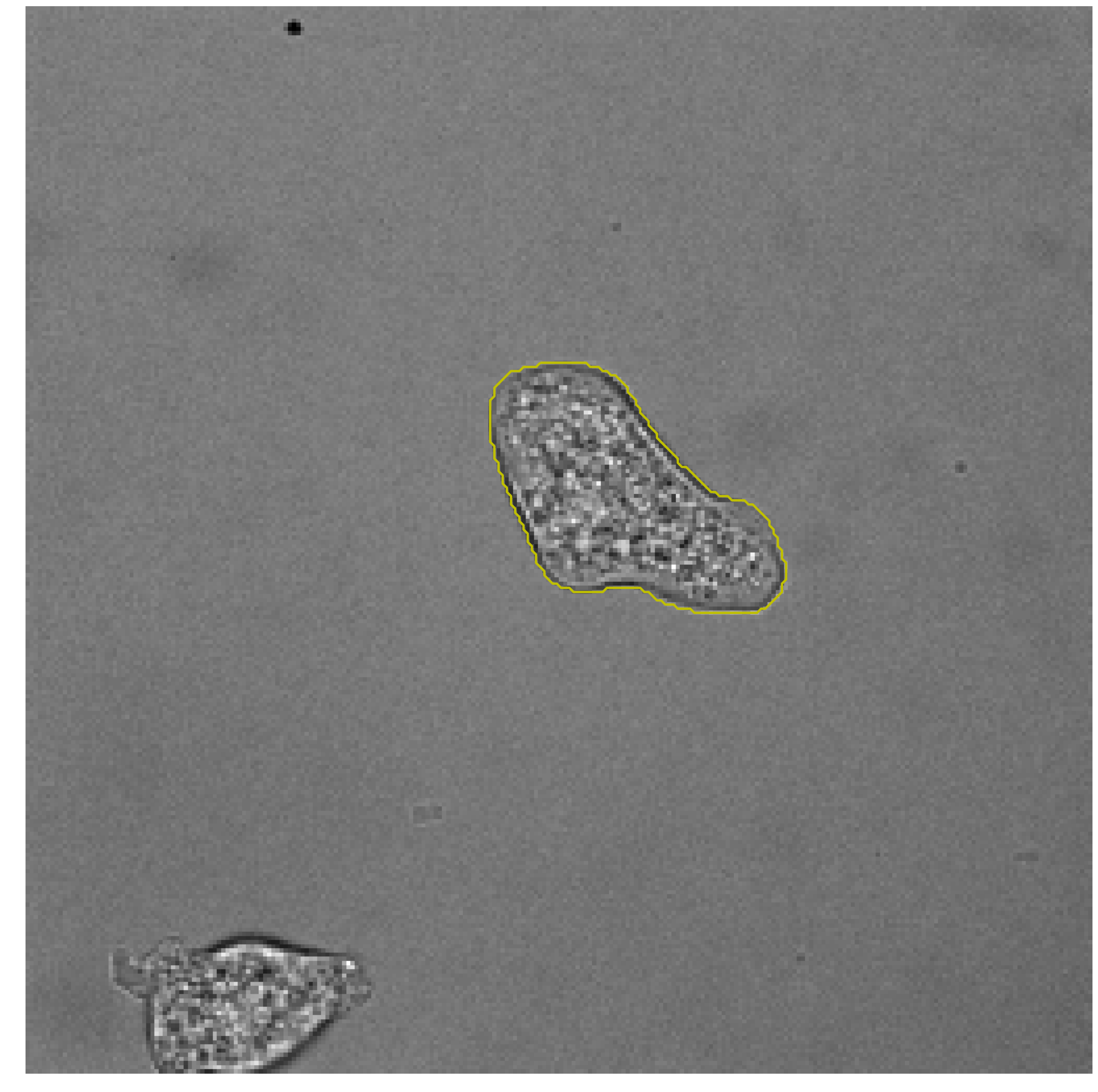}&
			\includegraphics[height=0.20\linewidth]{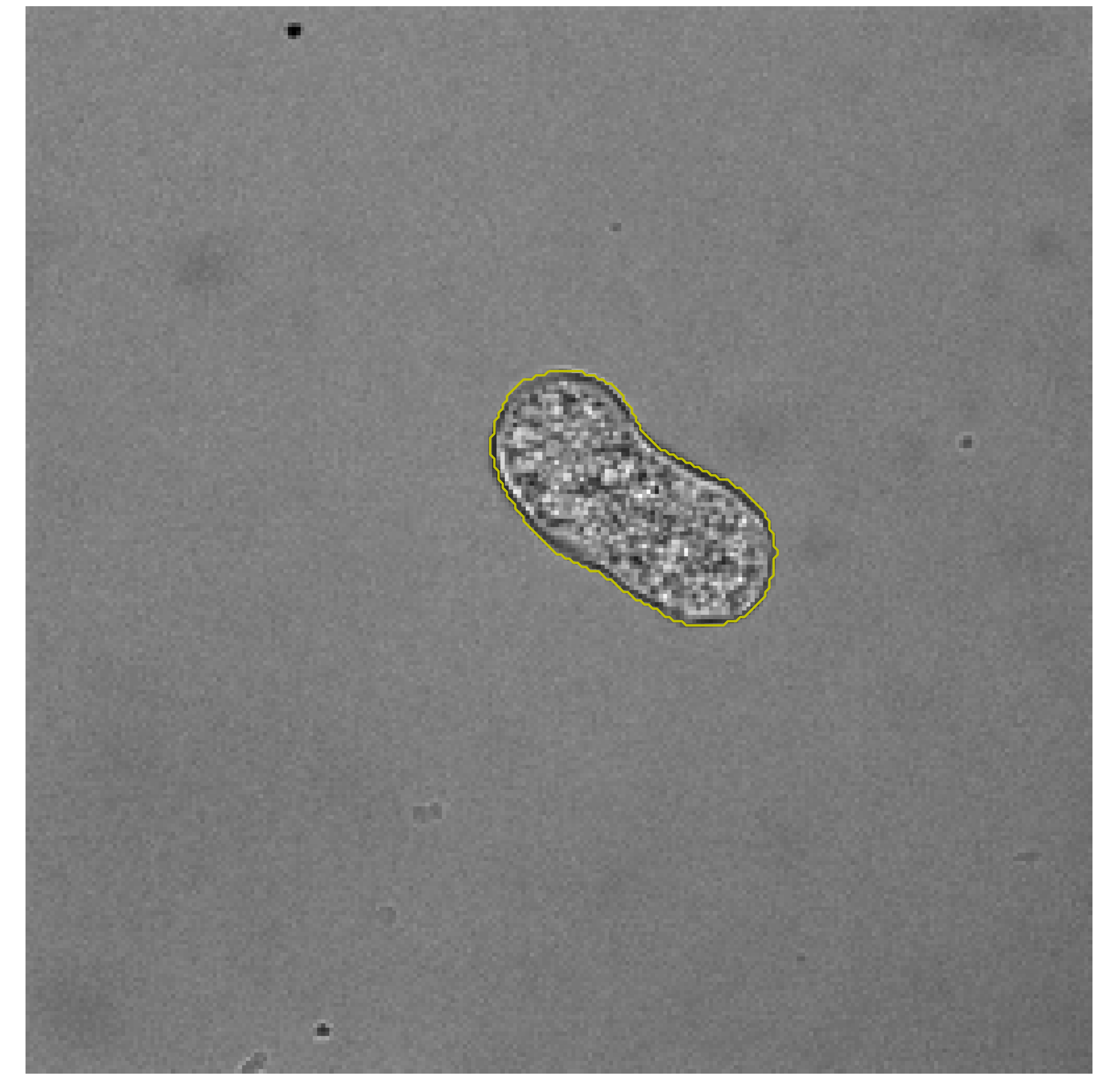}
			\\
			\includegraphics[height=0.20\linewidth]{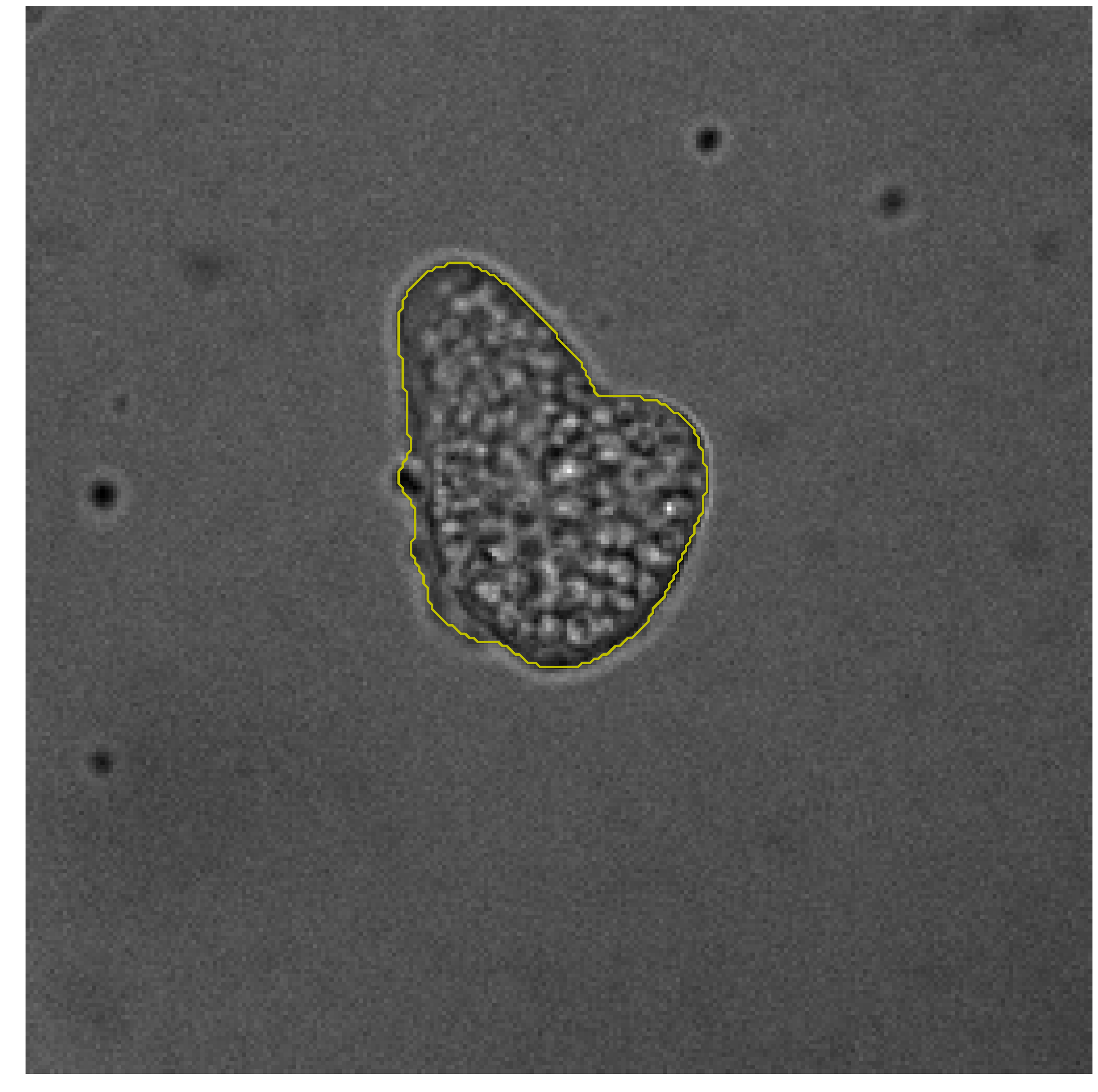}&
			\includegraphics[height=0.20\linewidth]{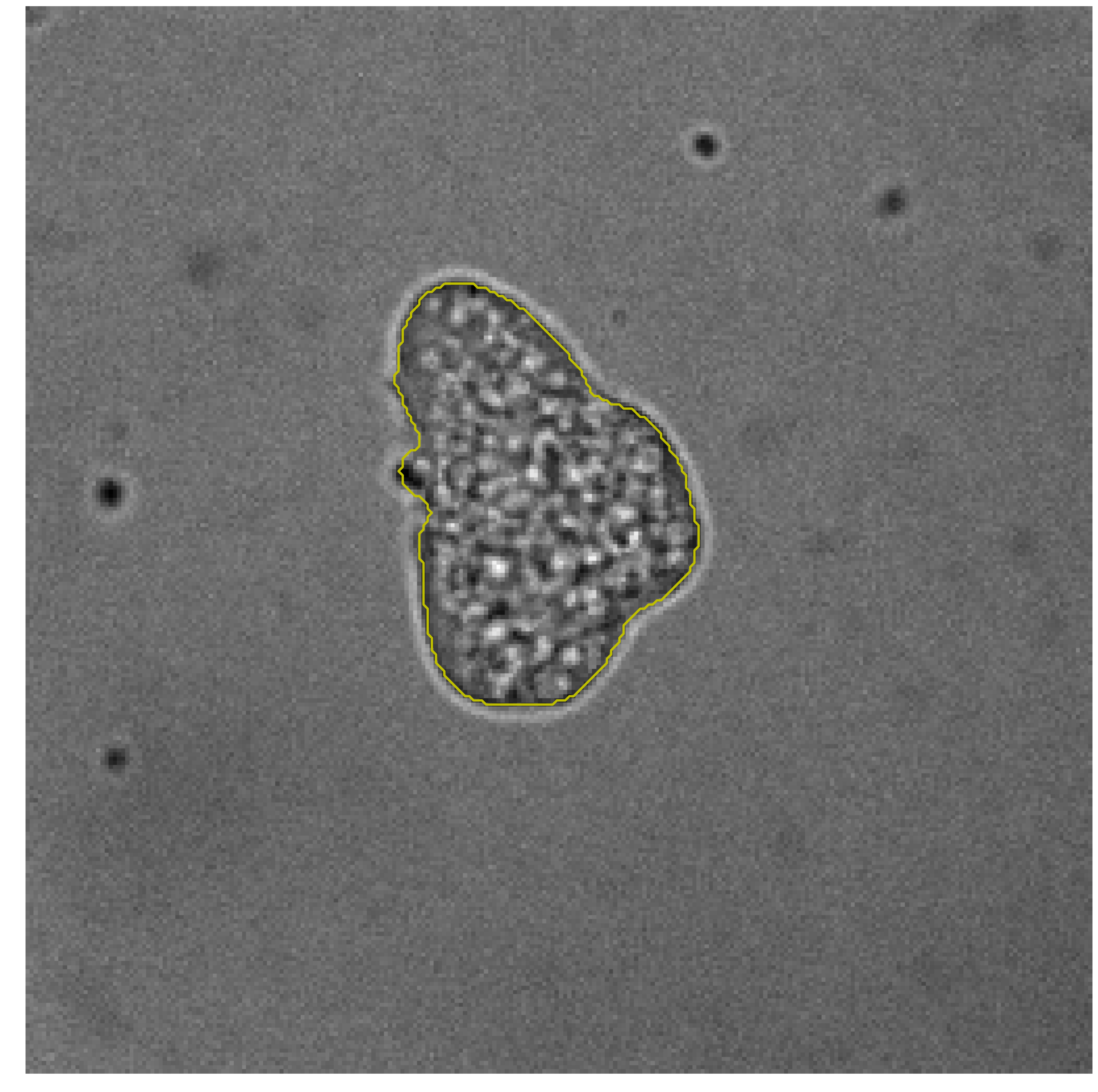}&
			\includegraphics[height=0.20\linewidth]{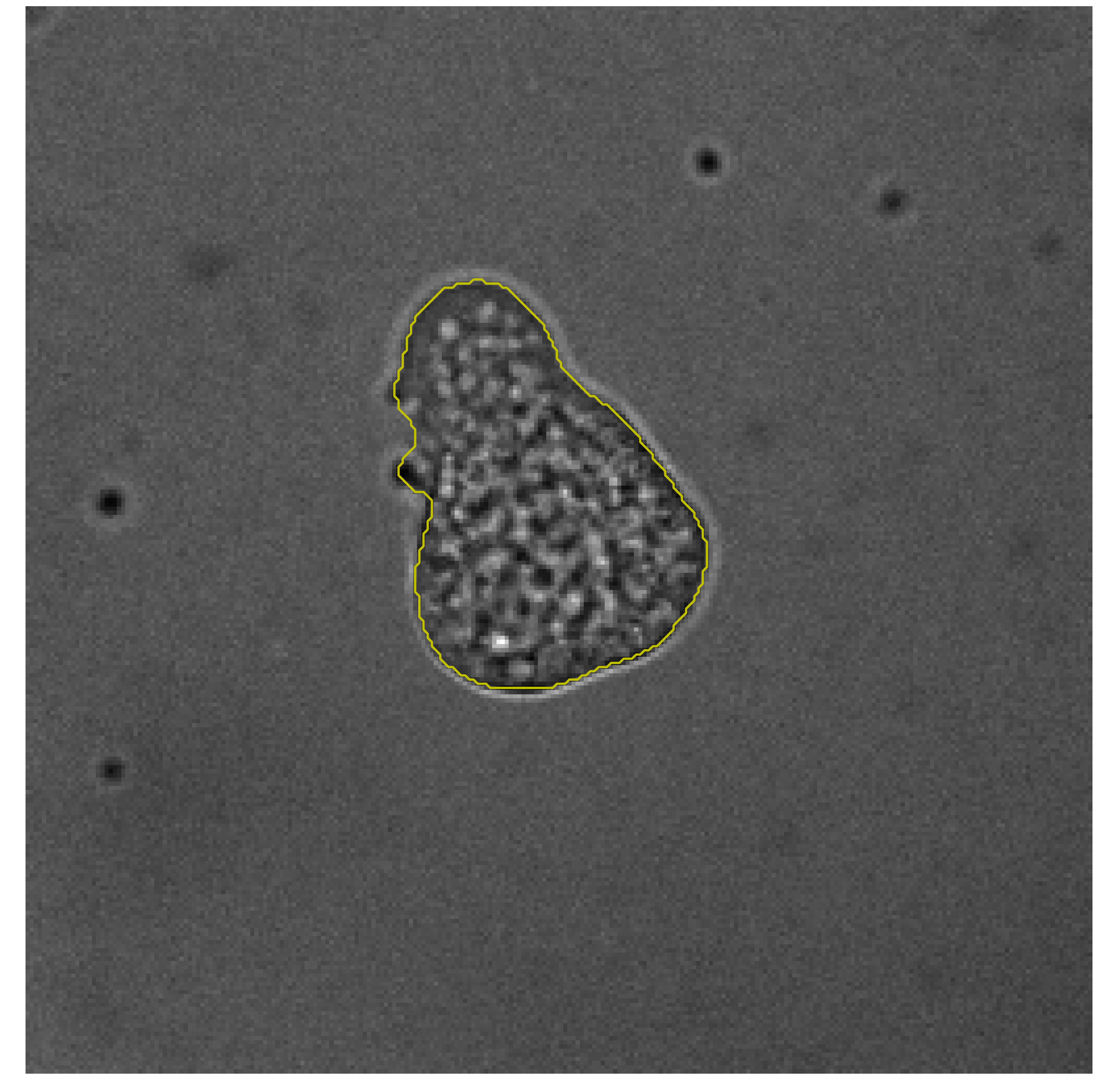}&
			\includegraphics[height=0.20\linewidth]{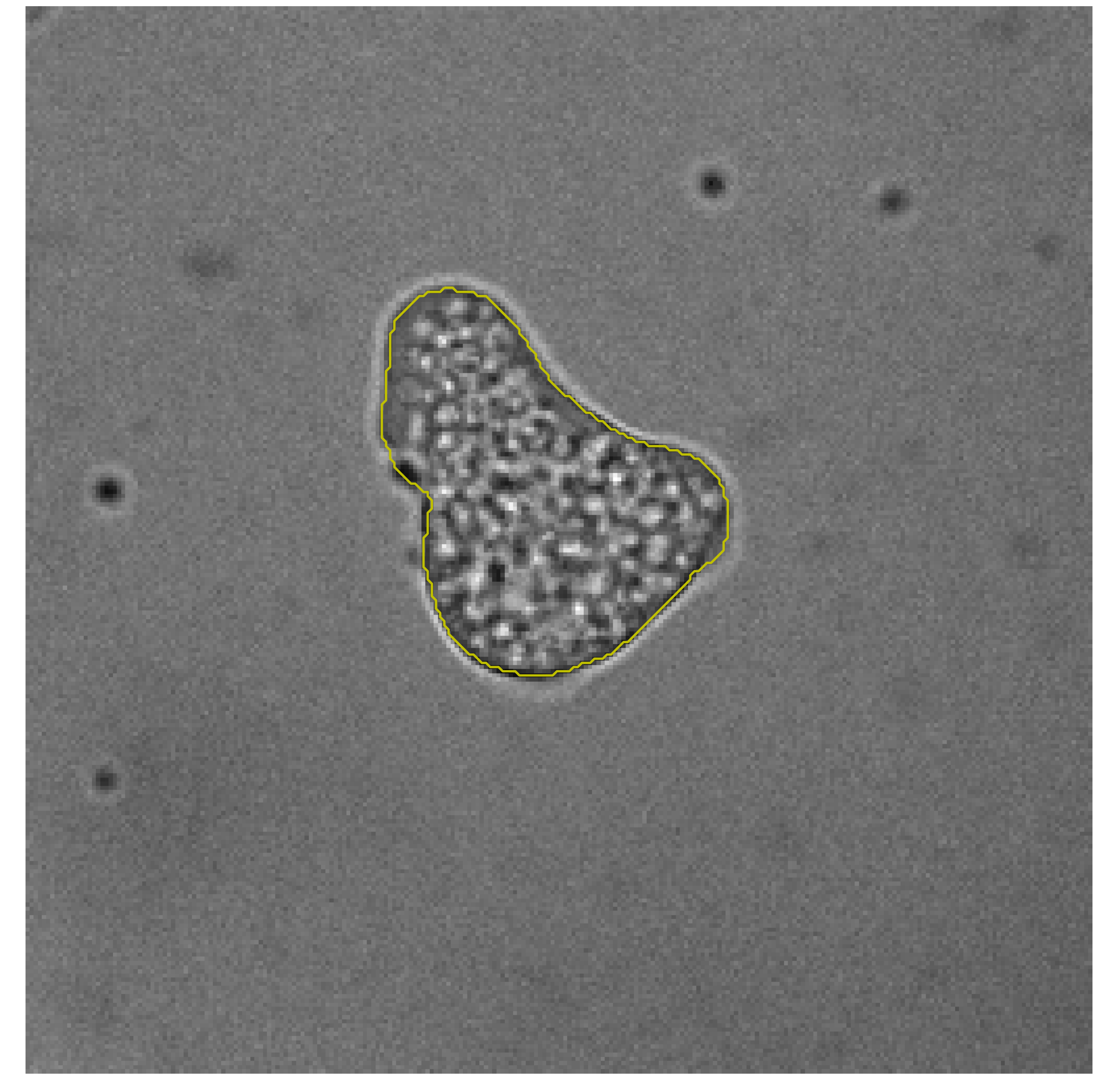}
			\\
			\includegraphics[height=0.20\linewidth]{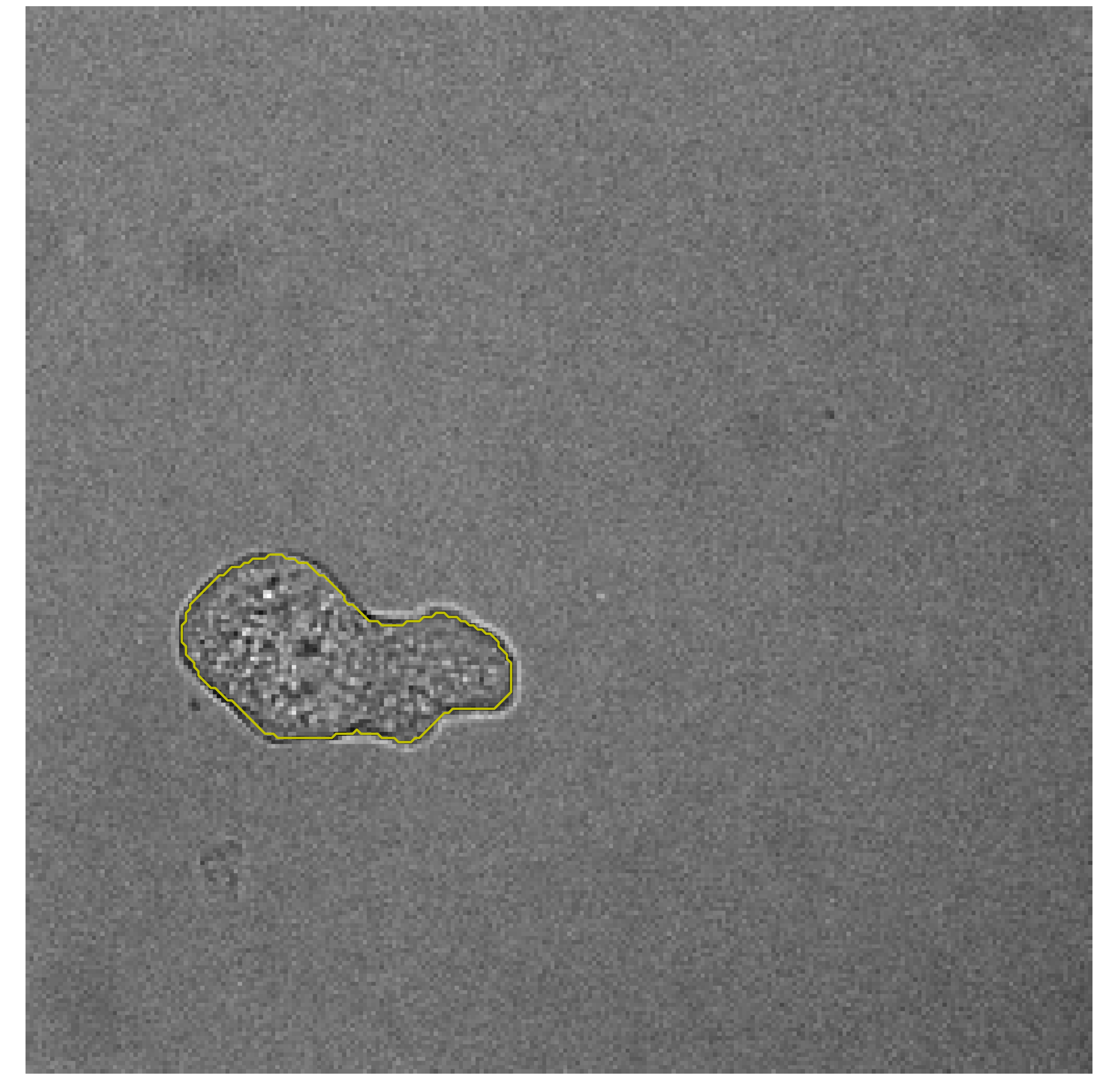}&
			\includegraphics[height=0.20\linewidth]{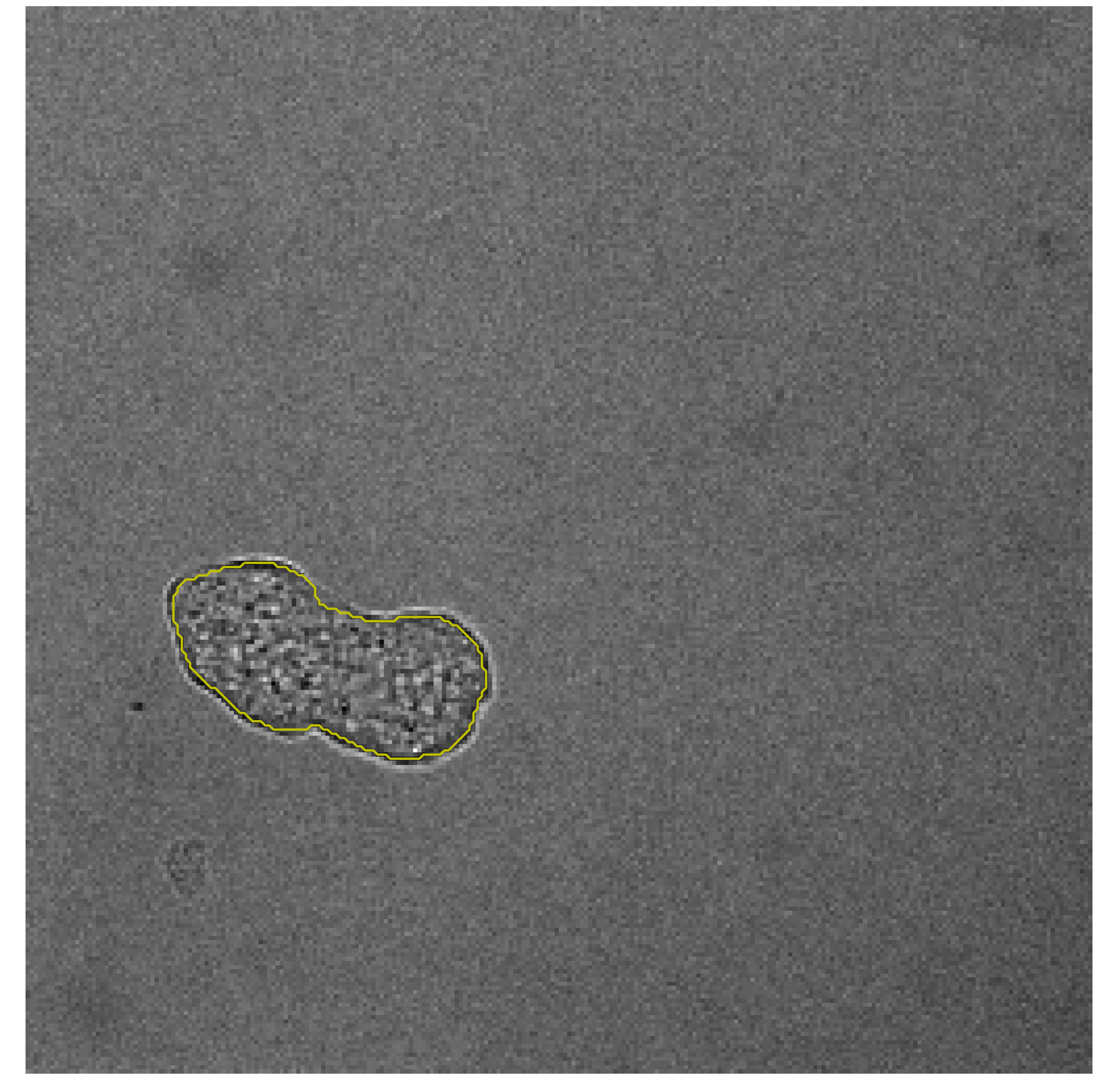}&
			\includegraphics[height=0.20\linewidth]{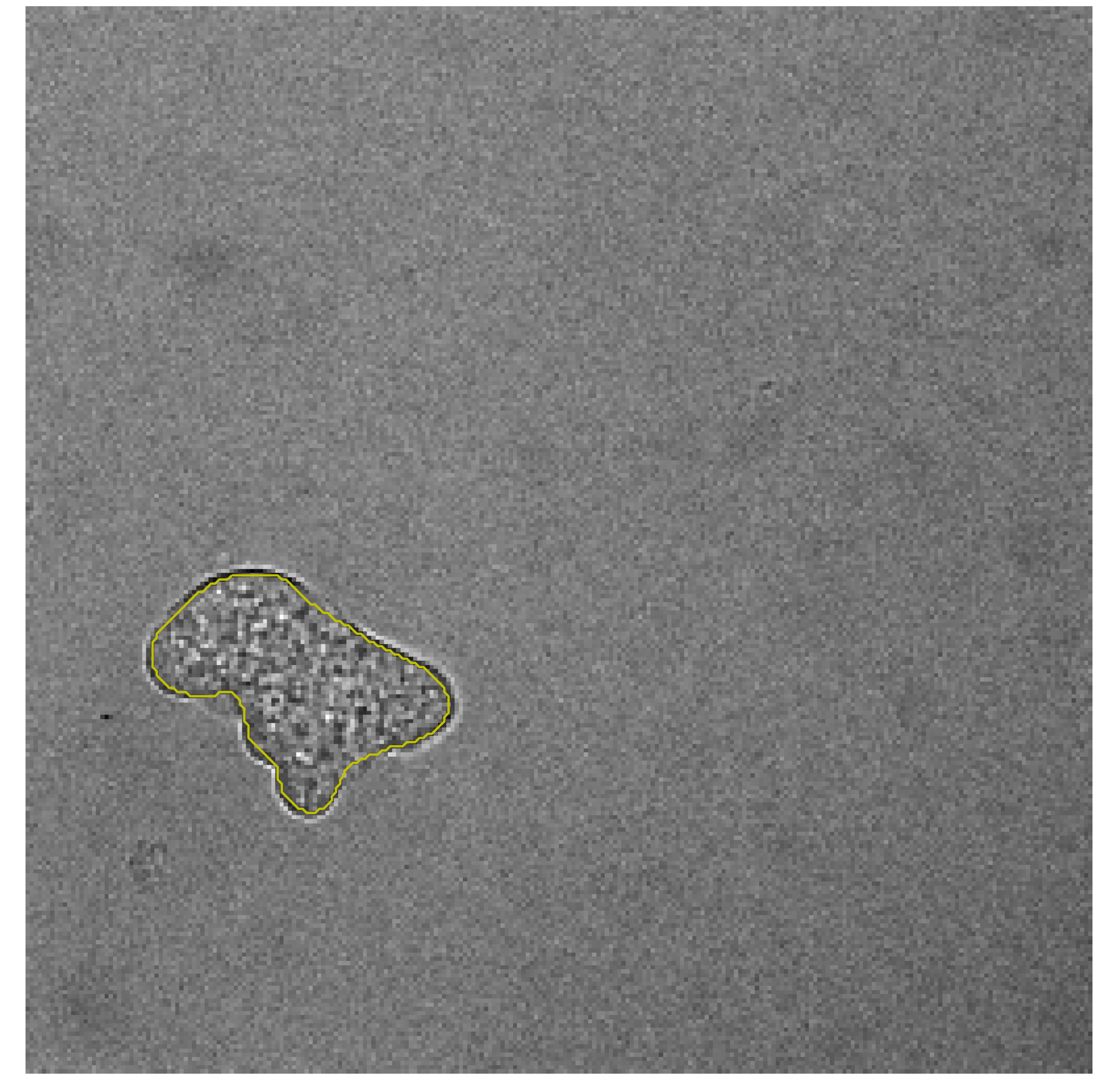}&
			\includegraphics[height=0.20\linewidth]{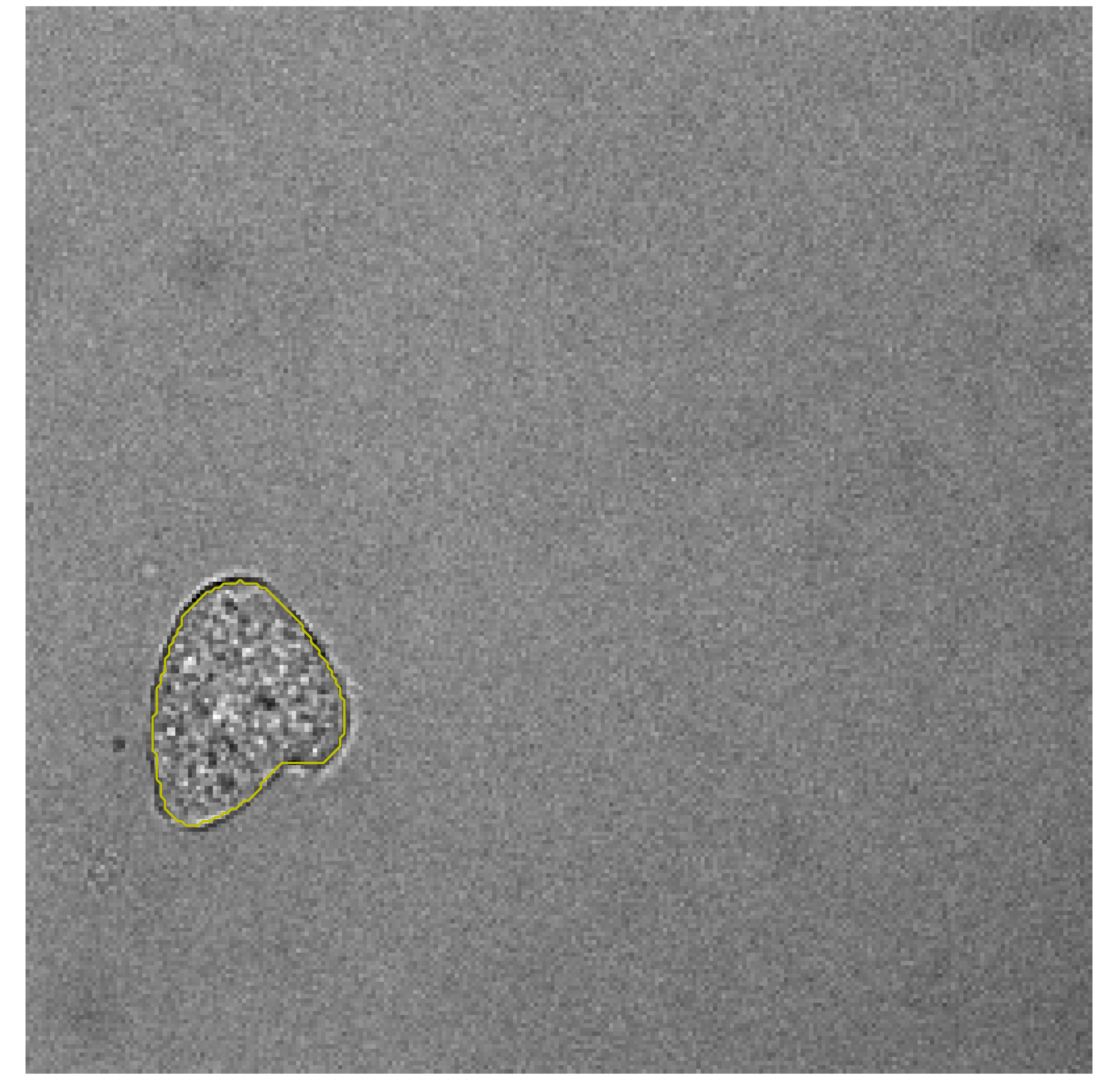}
			\end{tabular}
			\caption{\small Top: Image sequences with segmented cell contours in yellow for four sample sequences (from top to bottom) from Glass + Control, Glass + Inhibitor, Fibronectin + Control, Fibronectin + Inhibitor experiments.
				Bottom: examples of extracted shape sequences from 
				different experimental conditions.} \label{fig:segmentation}
			\vspace{-0.2in}
		\end{center}
		\vspace{-1pt}
	\end{figure}

	\subsection{Related Works}
	There exists an extensive literature on quantifying  kinematics, specifically trajectory analysis, for single-cell motility 
	\cite{campos2010persistent, miyoshi2003characteristics}. A few papers have also investigated the dynamics of cell migration, albeit from a biological/biophysical perspective \cite{boquet2017bioflow, holmes2012comparison, moreno2020modeling, camley2017crawling}. The current biophysical models for cell migration apply only to specific cell types as the models differ drastically from one cell type to another~\cite{ moreno2020modeling, camley2017crawling}. Further, such modeling is geared towards simulations and assessments of physical properties in cell migration, especially kinematics.
	Analyzing and characterizing cell membrane protrusions  \cite{tsygankov2014cellgeo,collier2017image, ducroz2012characterization, driscoll2012cell}, which demonstrate wave-like patterns and guide amoeboid motility, is another common research direction. 
	Tweedy et al. \cite{tweedy2013distinct} analyzed different motility modes exhibited by a migrating cell using principal components of Fourier shape descriptors, with the latter computed from individual shapes. In~\cite{dufour2014signal}, the authors employed various global shape features, 
	to classify and interpret differences between amoeba populations. In contrast, the analysis of local morphodynamics of cell membrane is based on spectral decomposition of spatiotemporal features  \cite{ma2018profiling} obtained by dividing cell membrane into local sectors. 
	However, these works do not account for the global temporal correlations in shape evolution, a crucial aspect of amoeboid motion. This deficiency motivates development of sophisticated mathematical representations and statistical models of dynamic shape evolution, which is not restricted to a specific type of analysis, but can be adopted to different avenues in cell biology as described in the following sections.
	
	In terms of shape analysis, there exists a large literature on techniques for shape analysis of planar contours. While older techniques represented contours using labeled points, or landmarks, the recent focus has been on parametrized curves. The use of elastic shape analysis of parametrized curves \cite{younes-elastic-distance,younes-michor-mumford-shah:08,srivastava_etal_PAMI:11,joshi-klassen-cvpr:07,BauerAppl2} has gained attention due to superior mathematical properties and good practical performance. A large majority of elastic shape analysis literature focuses on static shapes, with scant attention paid to analyzing time-varying shapes. There are some exceptions however. For example, using Dryden {\it et al.} have study dynamical shapes, albeit using Kendall's shape representations, in several contexts~\cite{Kume:2007,Kenobi:2010}. Also, the papers on activity recognition have studied dynamical shapes as trajectories in shape spaces \cite{abdelkader-etal-CVIU:11}. The use of time series models to analyze shape sequences of cells is a novel approach pursued in the current paper. 

	\subsection{Our Approach}
	
	Focusing attention on shapes of cell boundaries, we will utilize tools from {\it elastic shape analysis} of curves to characterize cell motility.  The main challenges in modeling shape dynamics are the infinite-dimensionality and nonlinearity of shape spaces under elastic representations. While time-series analysis methods for finite-dimensional Euclidean domains are well established, the corresponding solutions for shape spaces are relatively few. We will develop an approach that uses: (1) the geometry of shape spaces to flatten nonlinear representations of cell shapes into Euclidean variables and (2) principal component analysis (PCA) to reach finite dimensionality. Specifically, we will use a combination of inverse exponential maps and parallel translations on shape spaces to represent a shape sequence by a Euclidean time series. Then, we will model this Euclidean series using classical VAR and GARCH models and perform parameter estimation. We will use these estimated parameters to analyze and compare shape dynamics across appropriately defined classes. This work aims to develop representations that can capture essential dynamics of membranous shapes and develop statistical tools for related biomedical inferences. Such tasks include simulation of motile cellular sequences, shape prediction, clustering, and classifying predominant motility patterns in a cell population. A pictorial overview of this pipeline is shown in Fig.~\ref{fig:pipeline}


	In a preliminary paper \cite{deng-ICIP:2020}, we developed some parts of this pipeline. We showed that nonlinear representations of cell shapes can be modeled as a Euclidean time series by exploiting the differential geometry of shape spaces.
	In particular, we demonstrated the efficiency of the model in shape reconstruction and prediction of cellular shapes. The current paper presents extends to different time-series models and parameter estimation schemes, which can be employed over the Euclidean representations of shape sequences. The estimated parameters are then employed for shape-time series classification of amoeboid motility modes demonstrated by  \textit{E. histolytica} when subject to different biological experimental conditions. 
	
	The main contributions of this paper are as follows:
	\begin{enumerate}
		\item It uses the differential geometry of the shape space of planar, elastic curves, followed by PCA-based dimension reduction,  to represent shape sequences as low-dimensional, Euclidean time series. 
		\item It applies and fits classical time-series models, such as the vector autoregressive (VAR) \cite{ziegel1995tie-var} model and DCC-GARCH model \cite{Orskurg-dccgarch}, on the resulting time series and pursues VAR models for subsequent statistical analyses. 
		\item It validates the effectiveness of VAR models by synthesizing complete shape sequences and by performing predictions of future amoeboid shapes from past data in time series.
		\item Finally, it utilizes VAR parameters to characterize and classify modes of cell motility in different experimental conditions. It obtains a high classification performance in conjunction with machine learning, as demonstrated through various simulation and microscopy data. 
	
	\end{enumerate}
	These are the first results in modeling and classifying cell motility using shape analysis, as most previous results relied only on cell kinematics for such classifications. We combine shape analysis with kinematics data to obtain superior results. This pipeline provides a promising direction of research for understanding migration patterns during cell utility.   
	
\subsection{Data Description}
As mentioned, we adopt the unicellular \textit{E. histolytica} as the prototype to demonstrate the efficiency of the proposed method in analyzing cellular migration patterns. During a classical amoeboid movement (e.g., \textit{E. histolytica} on glass), the actomyosin contraction \cite{boquet2017bioflow} enhances intracellular pressure allowing bleb formation and induces contraction at the cell rear. One of the enzymes (in the mammalian cell) involved in myosin contraction is ROCK (Rho-dependent kinase)\cite{julian2014rho}. The importance of ROCK in podosome and invadosome  dynamics has been described in previous works \cite{van2019probing}. To analyze whether ROCK is involved in amoeboid migration on glass and fibronectin, we added a ROCK inhibitor ($100$ mM) during migration assays. In these experiments, the amoeba was seeded on 35 mm glass-bottomed or Fibronectin-coated imaging dishes.  In some experiments,  plated amoebae were incubated with a specific ROCK inhibitor (Y-27632) (BD Biosciences, USA). 
We recorded videos of cell movements with a spinning-disk confocal microscope (bright field mode). 
Thus, the data was obtained for four different biological conditions, where amoebas are on glass and fibronectin with or without rho-kinase inhibitor. 

The application and effectiveness of the proposed framework depend on the robust segmentation of individual cells from microscopy data. Although this segmentation is an interesting problem in itself, it is not the focus of this paper, which rather focuses on analysing the shape dynamics of the cells. We employed an existing approach \cite{sarkar2020multiamoebanet} designed explicitly for segmentation of amoeboid cells from brightfield microscopy to extract the cell boundaries. Sample images of the cell sequences captured using the brightfield mode overlaid with the segmented contours (in yellow) are shown in Fig.~\ref{fig:segmentation}. Our approach for dynamics shape modeling assumes that contour sequences extracted from microscopy videos are directly available to us.

\section{Mathematical Representation of Shape Dynamics}
We start by reviewing the elastic shape analysis of planar closed contours, and then apply it to derive Euclidean time-series representations of cell shape sequences.

	\subsection{Elastic Shape Analysis of Contours}
The main challenges in modeling and analyzing shape sequences as time series are the nonlinearity and infinite-dimensionality of shape spaces \cite{srivastava-klassen:2016, srivastava_etal_PAMI:11}. We will use parallel transport of the velocity vectors to linearize shape representations without significant distortions and use principal component analysis (PCA) to reduce dimensions. 
	
	To develop a mathematical representation of shape evolutions, we use the elastic shape analysis \cite{srivastava-klassen:2016, srivastava_etal_PAMI:11}. 
	In this theory, each closed parametrized curve $y: \s^1 \to \real^2$ is represented by its Square-Root Velocity Function (SRVF) $q: \s^1 \to \real^2$ given by: $q(t) =  {\dot{y}(t) \over \sqrt{| \dot{y}(t)|}}$.
	The use of SRVF greatly simplifies the shape analysis of curves, especially in imposing invariance to rotation, translation, scaling, and re-parameterization of $y$. Note that each of these transformations is considered shape-preserving transformations, and, therefore, one needs to be invariant to them in shape analysis. 
	
\begin{figure}
		\begin{center}
			\begin{tabular}{|c|}
				\hline
				\includegraphics[height=0.3in]{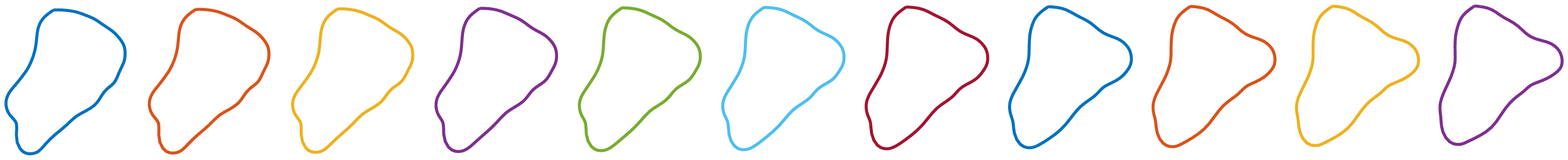}  \\
				\hline
				\includegraphics[height=0.3in]{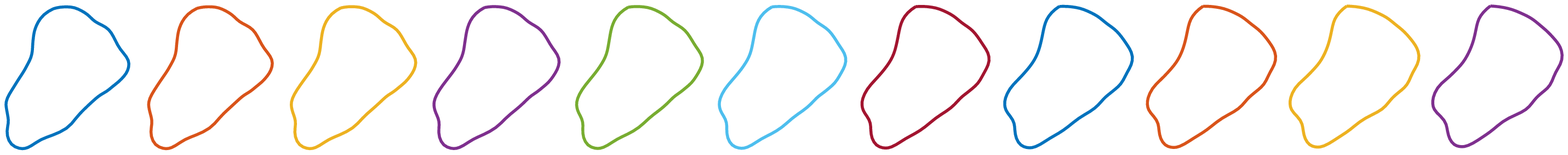}  \\
				\hline
				\includegraphics[height=0.3in]{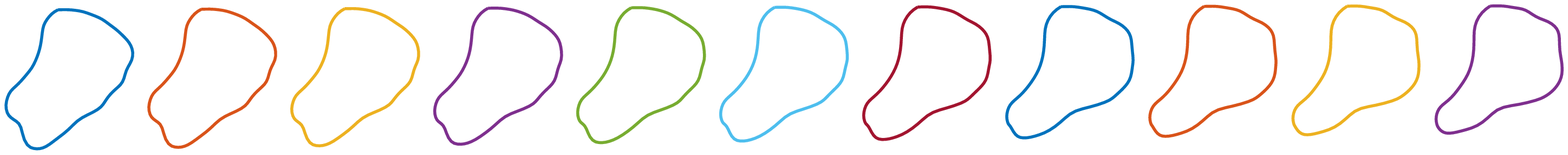}  \\
				\hline
			\end{tabular}
			\caption{\small Examples of geodesic paths between some cell shapes.
				Computations are done in SRVF shape space ${\cal S}$ but displayed as curves.} \label{fig2}
			\vspace{-0.2in}
		\end{center}
	\end{figure}

	The SRVF $q$ of a curve $y$ is already invariant to the translation of $y$. 
	If a curve $y$ is rescaled to have length one, then its SRVF $q$ has $\ltwo$ norm of one, i.e. 
	$q \in \s_{\infty}$, the unit Hilbert sphere. If $y$ is rotated by a matrix $O \in SO(2)$, and 
	re-parameterized by a diffeomorphism $\gamma$, then the new SRVF is given by 
	$O (q \circ \gamma) \sqrt{\dot{\gamma}}$. Since all of these preserve shape of $y$, 
	the shape of $y$ is represented by the equivalence class 
	$[q] = \{ O  (q \circ \gamma) \sqrt{\dot{\gamma}} | O \in SO(2), \gamma \in \Gamma\}$,
	that contains SRVFs of
	all possible rotations and re-parameterizations of $y$. The set of all shapes is 
	denoted by ${\cal S}$. ${\cal S}$ is an infinite-dimensional, nonlinear space and that limits
	our ability to perform statistical modeling and analysis of shapes. For instance, given any two 
	shapes, $[q_1]$ and $[q_2]$, we can not take a linear combination of these two shapes. 
	Using previously developed tools, 
	Fig.~\ref{fig2} show examples of geodesic paths between some static cell shapes. Each row shows a uniformly-spaced points along the geodesic between the first and the last shapes.  
	This geodesic computation can also be used to find the shooting vector. 
	That is, given two shapes $[q_1]$ and $[q_2]$, first we compute a geodesic path 
	$g: [0,1] \to {\cal S}$ such that $g(0) = [q_1]$ and $g(1)= [q_2]$. Then, the initial 
	velocity $\dot{g}(0)  \in T_{[q_1]}({\cal S})$ is also called the {\it shooting vector} 
	from $[q_1]$ to $[q_2]$. 
	These quantities -- geodesics, shape distances, and shooting vector -- are invariant to 
	rigid motions of curves, and are instrumental in separating cell kinematics from its morphology for our purposes. 
	
	So far we have a shape space ${\cal S}$ of planar curves. A shape sequence associated with a migrating cell corresponds to a discrete time-series on ${\cal S}$. 
	Next we develop Euclidean representations for such sequences. 
	
	\subsection{Transported Velocity Fields}
\begin{figure}
		\begin{center}
			\begin{tabular}{|c|c|c|}
				\hline
				\includegraphics[height=0.5in]{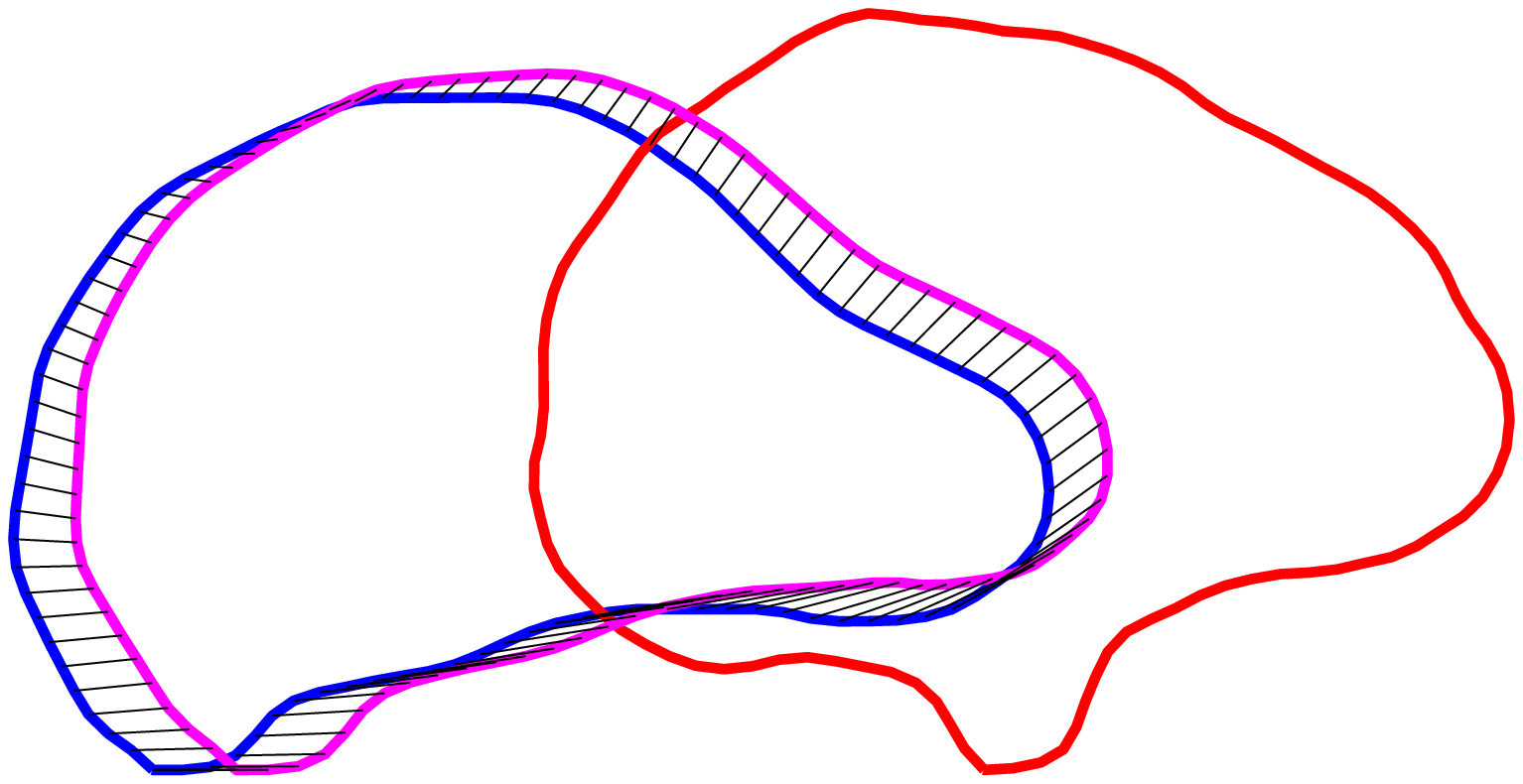}  &
				\includegraphics[height=0.5in]{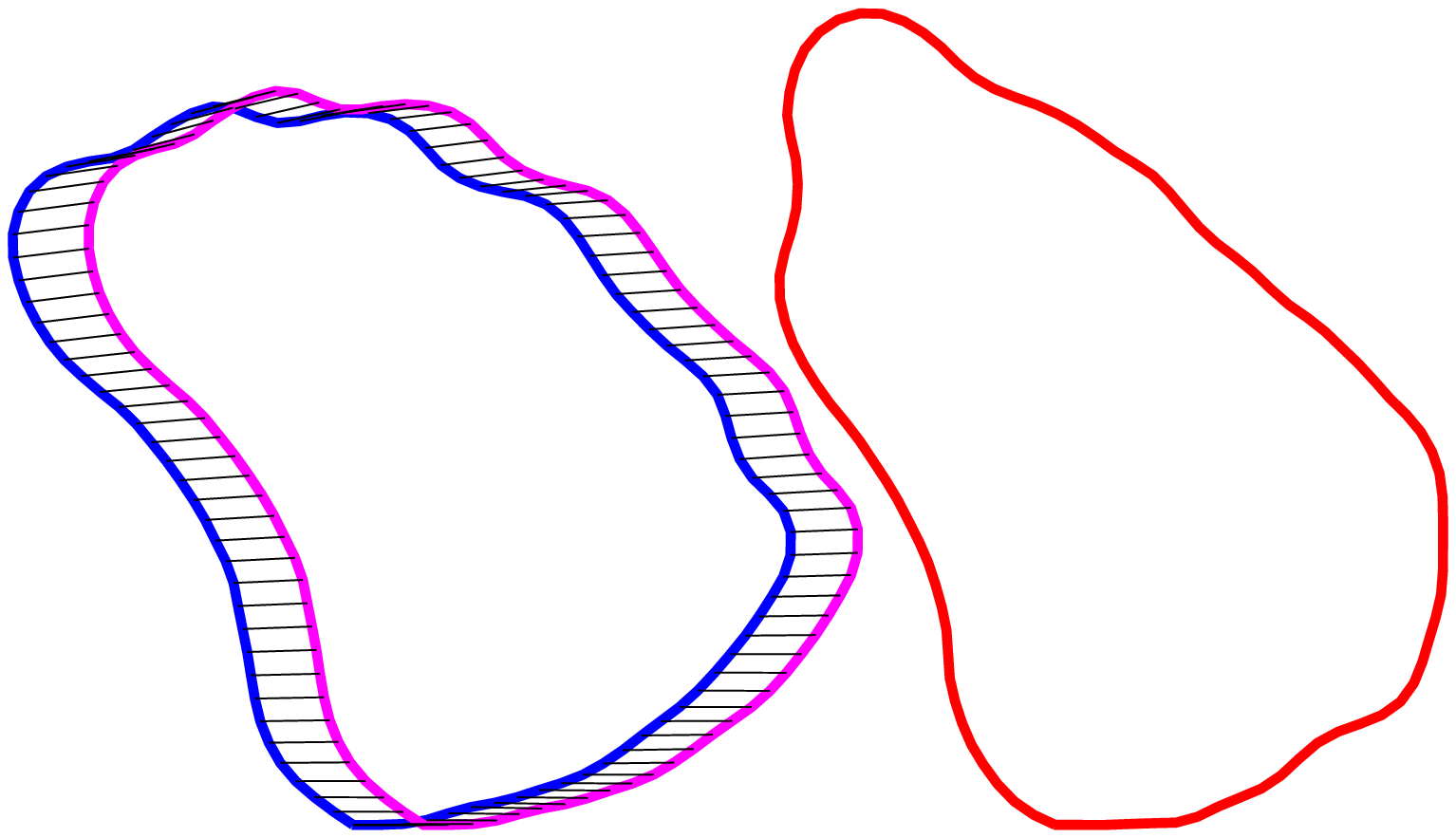}  &
				\includegraphics[height=0.5in]{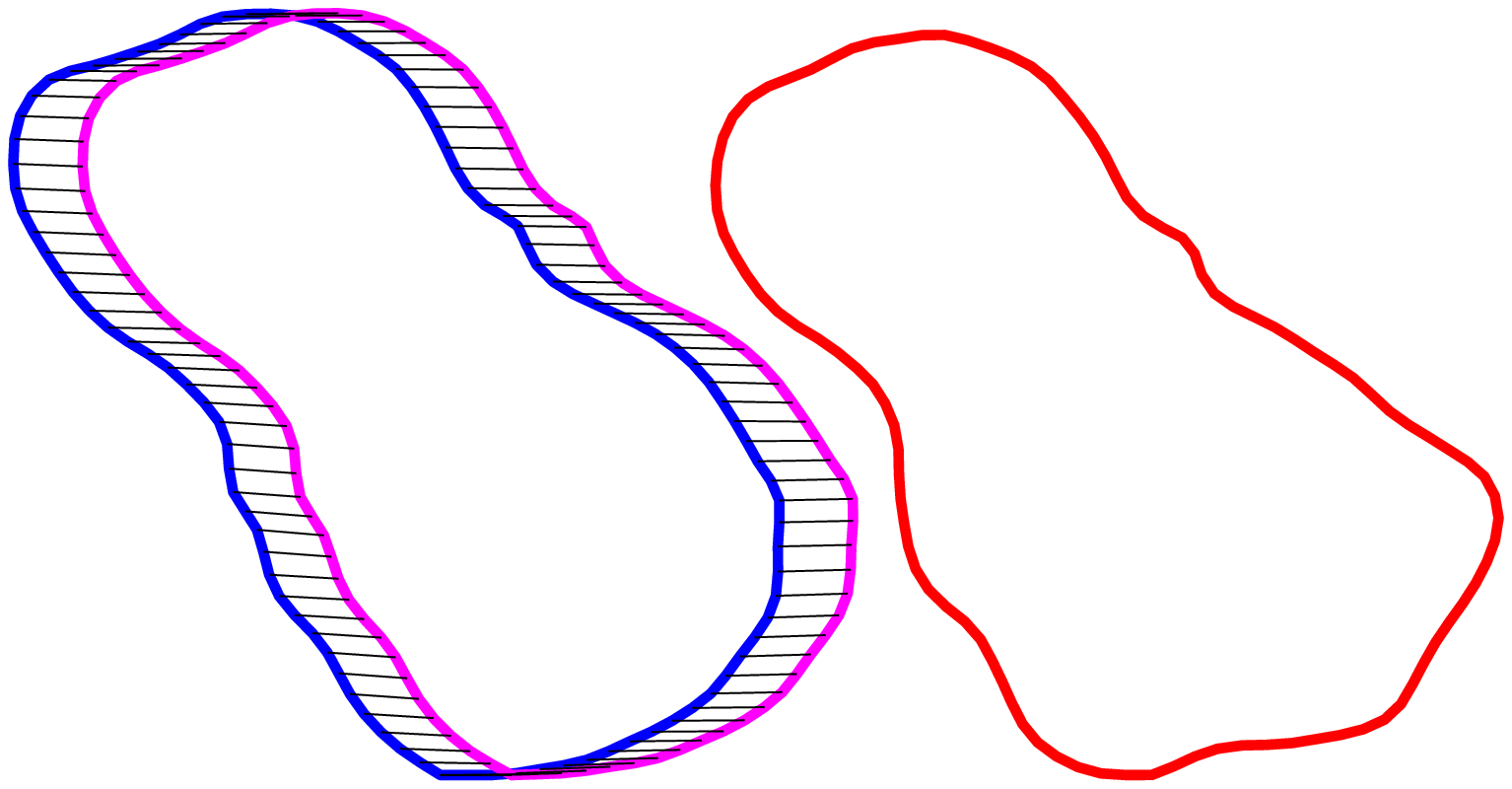}  \\
				\hline
			\end{tabular}
			\caption{\small Examples of computing shape velocities $\dot{\alpha}(\tau)$ in the shape space.
				As earlier, the computations are performed in ${\cal S}$ space but displayed 
				for convenience in the curve space.} \label{fig3}
			\vspace{-0.2in}
		\end{center}
	\end{figure}

    Let $\alpha: {\cal I} \to {\cal S}$ represent a discrete-time evolution of a shape over the observed times ${\cal I} = \{0, 1, 2, \dots, T-1\}$. The next question is: What is a useful mathematical representation for comparing 
	and modeling this process? As stated earlier, the difficulty in answering this question comes from the nonlinearity and infinite dimensionality of ${\cal S}$. So, we use the changes in the shapes, denoted by $\{\dot{\alpha}(\tau)\}$, instead of 
	shapes $\{\alpha(\tau)\}$ themselves, to represent the dynamics. Here $\dot{\alpha}(\tau) = \exp_{\alpha(\tau)}^{-1}(\alpha(\tau+1))$ stands for 
	the shooting (tangent) vector form $\alpha(\tau)$ to $\alpha({\tau+1})$. Note that the expressions for the exponential map and its inverse for the shape space ${\cal S}$ have been discussed in the book \cite{srivastava-klassen:2016}. This representation is able to deal with non-linearity in the data and the subsequent use of PCA help reach a finite representation space.
	Fig.~\ref{fig3} shows three examples of computing shooting vectors or shape velocities. In each panel, we show 
	two shapes and the infinitesimal deformation (shooting vector field) on the first shape that takes the first to the second. 
	However, these velocities 
	$\dot{\alpha}(\tau) \in T_{\alpha(\tau)}({\cal S})$ are elements of different tangent spaces at different times, and we need to 
	bring them to the same space for comparisons. This is accomplished using parallel transport on the shape space ${\cal S}$. 
	
	\vspace{-1pt}
	
	\begin{definition}
	({\bf Transported Square-Root Velocity Field or TSRVF}) \cite{zhang-etal-covariance}:
		For a shape sequence $\alpha: {\cal I} \to {\cal S}$, we define its  transported square-root velocity field (TSRVF) 
		according to: 
		\begin{eqnarray}
			F_{\alpha}(\tau) = (\dot{\alpha}(\tau))_{\alpha(\tau) \rightarrow \alpha(0)} \in T_{\alpha(0)}({\cal S}),\ \tau = 1,\dots, T-1\ ,
		\end{eqnarray}
		where the subscript $\alpha(\tau) \rightarrow \alpha(0)$ denotes the parallel transport of 
		$\dot{\alpha}(\tau)$ from $\alpha(\tau)$ to $\alpha(0)$ along the path $\alpha$. 
		Similarly, define the integrated TSRVF (I-TSRVF) to be:  $
			H_{\alpha}(\tau) = \sum_{s=0}^{\tau} F_{\alpha}(s)$.
	\end{definition}
	A shape sequence can be represented uniquely either by the pair $(\alpha(0), F_{\alpha})$ or 
	$(\alpha(0), H_{\alpha})$ depending on the application. That is, 
	given any one of these pairs, one can reconstruct $\alpha$ using covariant integration and parallel 
	translation. For instance, given $(\alpha(0), F_{\alpha})$, we can compute: 
	$$
	\alpha(\tau+1) = \exp_{\alpha(\tau)}((F_{\alpha}(\tau))_{\alpha(0) \rightarrow \alpha(\tau)})\ , \tau=1,2,\dots, T-2\ .
	$$
	The TSRVF $F_{\alpha}$ and I-TSRVF $H_{\alpha}$  are 
	time series in a vector space $T_{\alpha(0)}({\cal S})$ and 
	can be used for statistical modeling and analysis. Similar quantities, albeit for Kendall's shape analysis, have been utilized for studying shape dynamics in \cite{Kume:2007,Kenobi:2010} and related works.

\begin{figure}
		\centering
		\begin{tabular}{|c|c|}
			\hline
			$\alpha$ & \includegraphics[height=0.3in]{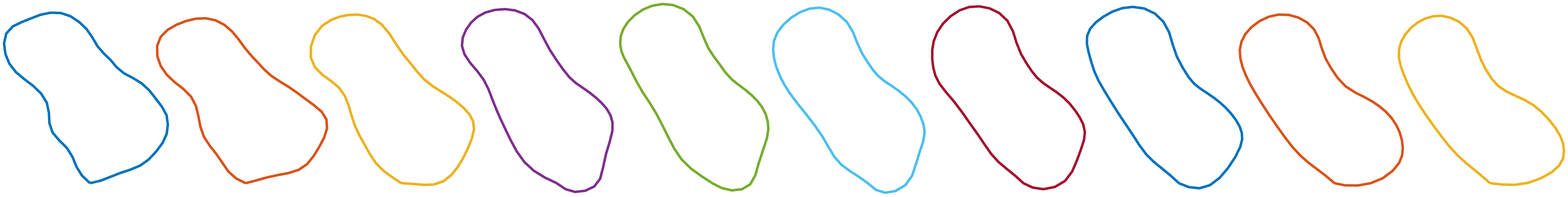} \\
			\hline
			$\hat{\alpha}$, $d = 1$ & \includegraphics[height=0.3in]{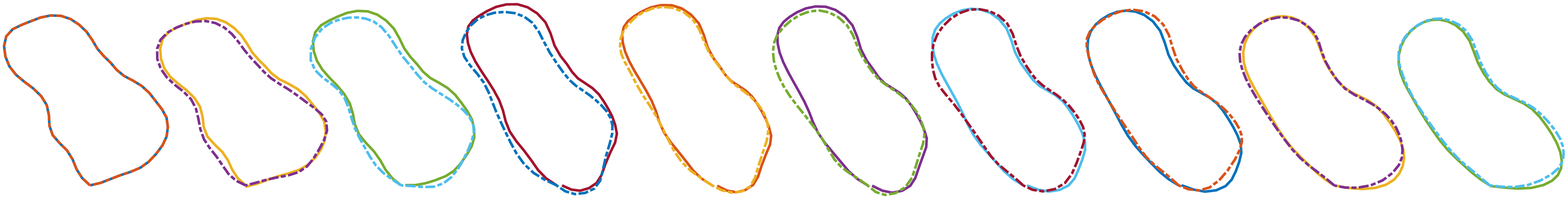} \\
			\hline
			$\hat{\alpha}$, $d = 5$ & \includegraphics[height=0.3in]{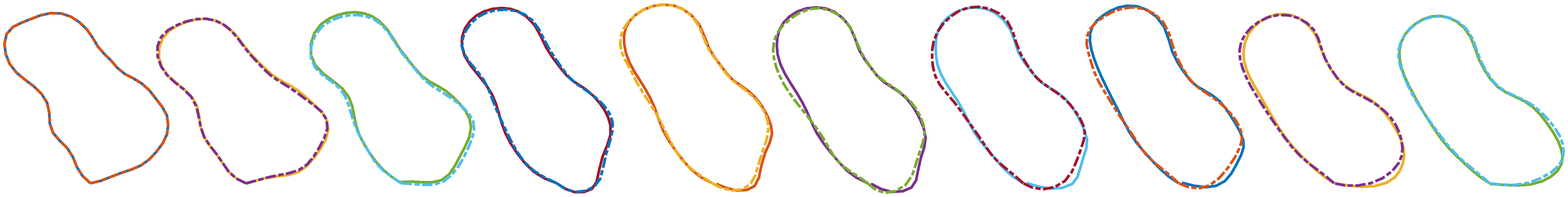} \\
			\hline
			$\hat{\alpha}$, $d = 10$ & \includegraphics[height=0.3in]{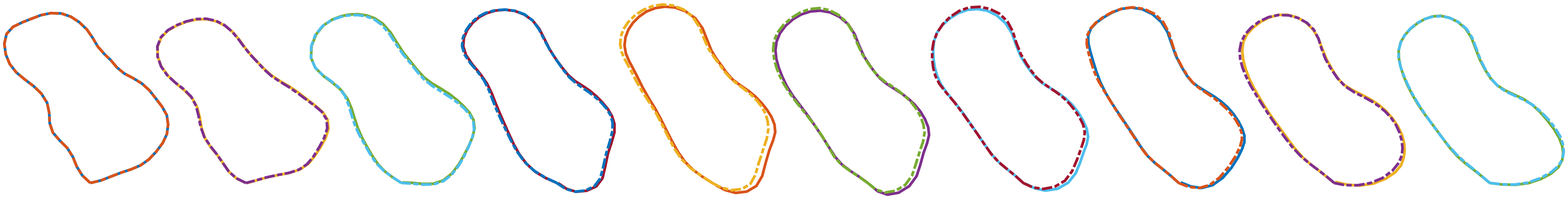} \\
			\hline
		\end{tabular}
		\begin{tabular}{c}	
			\includegraphics[height=1.in, width = 1.9in]{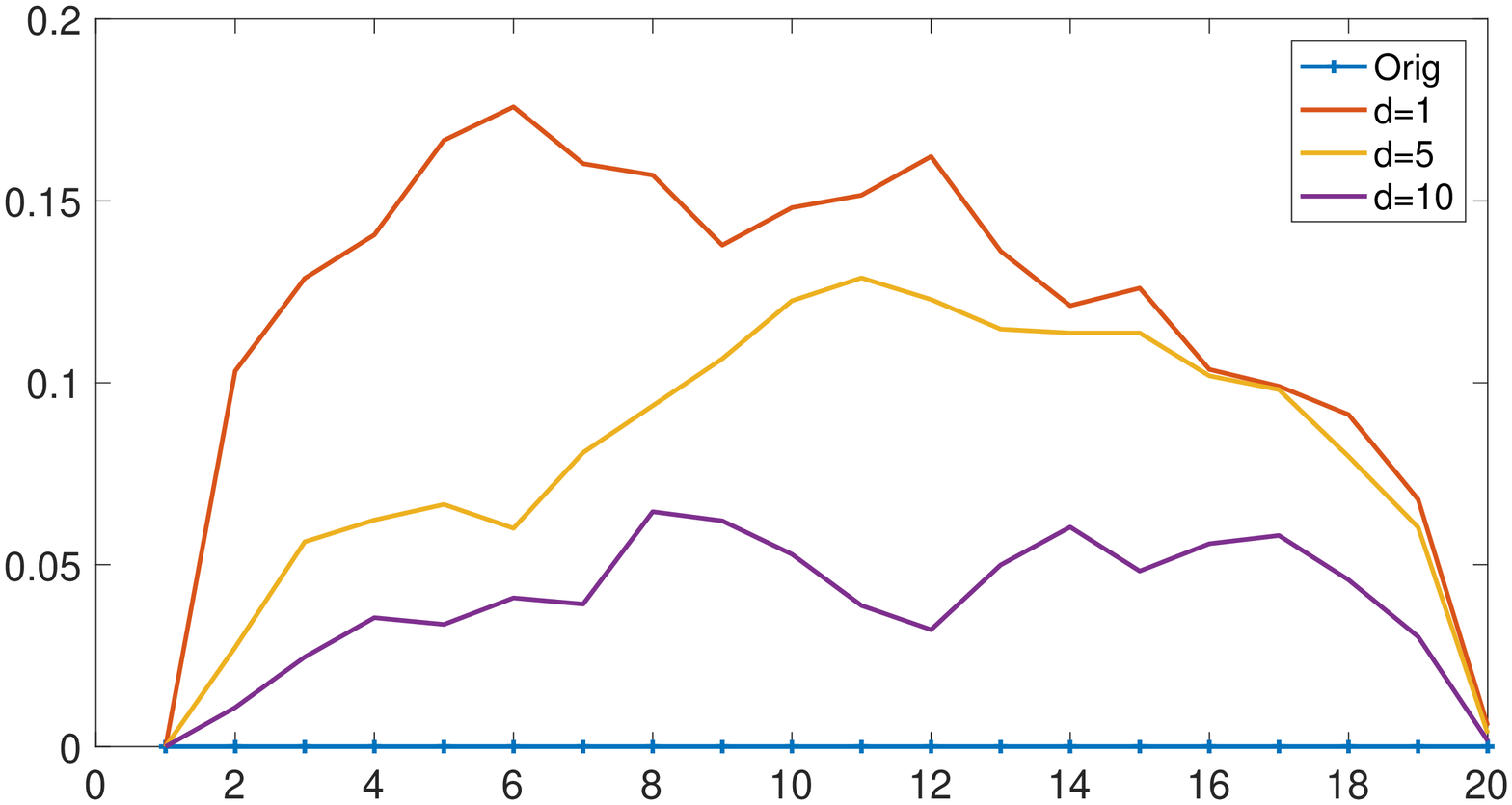} 
		\end{tabular}
		\caption{Reconstruction of a shape sequence $\alpha$ from its Euclidean representation TSRVF-PCA $x_{\alpha}$ for different values of $d$. The first rows show the original sequence $\alpha$ and the next three show the the reconstructed sequences for different $d$.The bottom plot quantifies the reconstruction error plotted versus $\tau$.} \label{fig:reconstruct} 
	\end{figure}

	Since the analysis remains the same whether we use $F_{\alpha}$ or $H_{\alpha}$, we will henceforth focus only on TSRVF.
	The next issue is that $F_{\alpha}$ are still infinite dimensional. 
	Let $\Pi: T_{\alpha(0)}({\cal S}) \to \real^d$ denote a linear projection to a finite-dimensional 
	subspace $\real^d$ of the tangent space $T_{\alpha(0)}({\cal S})$. The projection $\Pi$ in this paper will come from the principal component analysis (PCA) of the observed TSRVFs, as elements of $T_{\alpha(0)}({\cal S})$, while ignoring their time labels. 
	Let $x_{\alpha}(\tau) = \Pi(F_{\alpha}(\tau))$ denote the first $d$ PCA coefficients of $F_{\alpha}(\tau)$, forming a $d$-dimensional, Euclidean time series.  We will call $x_{\alpha}$ the TSRVF-PCA of the shape sequence $\alpha$.
	The forward mapping from the original shape sequence $\alpha$ to the representation $x_{\alpha}$ is given by: 
	\begin{eqnarray}
		\alpha \in {\cal S}^{\cal I}  \rightarrow  \dot{\alpha} \in (T{\cal S})^{\cal I}
		\rightarrow  F_{\alpha} \in (T_{\alpha(0)}({\cal S}))^{\cal I} \stackrel{\Pi}{\longrightarrow} x_{\alpha} \in (\real^d)^{\cal I}\ . \nonumber
	\end{eqnarray}
	Here $T{\cal S}$ denotes the tangent bundle of the shape space ${\cal S}$.
	Using appropriate constraints, one can invert $\Pi$ for use in mapping the Euclidean time series $x_{\alpha}$
	back to a shape sequence $\hat{\alpha}$, that approximates $\alpha$. In the PCA context, these constraints equate to setting the principal coefficients beyond the first $d$ to be zero. Fig.~\ref{fig:reconstruct} shows an example of reconstructing a shape sequence $\alpha$ from its representation $(\alpha(0), F_{\alpha})$ for different values of $d$. 
	The bottom plot shows the 
	shape wise reconstruction error $d_s(\alpha(\tau), \hat{\alpha}_d(\tau))$ versus $\tau$. We see that the reconstruction error is minimal for $d = 10$, 
	and very high for $d = 1$. The value of $d = 5$ seems to provide a good balance between the representation size and the reconstruction error.

	\section{Time-Series Models in TSRVF-PCA Space}
	Now we have $d$-dimensional Euclidean time-series $x_{\alpha}: {\cal I} \to \real^d$ associated
	with a shape sequence $\alpha: {\cal I} \to {\cal S}$. We will impose a statistical model on this time series and use this model to analyze shape dynamics of amoeba videos.
	For this purpose, we can use some existing 
	ideas involving auto-regression and moving averages~\cite{lutkepohl:book}
	
	\subsection{Common Time-Series Models}
	We will explore the VAR and GARCH models for modeling the $d$-dimensional time series $x_{\alpha}$. These are two most commonly used models in statistical time series analysis. 
	\begin{itemize}
		\item[I.] {\bf Vector Auto-Regressive or VAR Model}:
		A popularly-used model for Euclidean time series is VAR($p$), given by~\cite{lutkepohl:book}: 
		\begin{equation}
			x_{\alpha}(\tau) = c + \sum_{j=1}^p A_j x_{\alpha}(\tau-j)  + \epsilon_{\alpha}(\tau), \tau \in {\cal I}\ .
			\label{eqn:var-model}
		\end{equation}
		where: $p$ is the model lag, $c \in \real^d$ is a constant, $A_j \in \real^{d \times d}$ are coefficient matrices, and  $\epsilon_{\alpha}(\tau) \in \real^d$ is the observation noise, modeled as {\it i.i.d} 
		multivariate normal with mean zero and covariance $\Sigma$.
		Given an observed sequence $x_{\alpha}$ one can estimate model parameters $\Theta = (c,\ \{A_j\}, \Sigma)$
		using the generalized least square error
		criterion, as described later. 
		
		\item[II.] {\bf Dynamic Conditional Correlation - GARCH or DCC-GARCH Model}:
		Another popular time-series model is the multivariate  
		Generalized Autoregressive Conditional Heteroscedasticity or GARCH  model, with dynamic conditional correlation~\cite{Engle2002Dynamic} ~\cite{Engle2001theoretical}. Here we model:
		\begin{equation}
			\label{eqn: garch}
			\begin{split}				
				x_{\alpha}(\tau)  &= \mu(\tau) + e(\tau), e(\tau)= K^{1/2}(\tau) e(\tau), \\
				K(\tau) &= D(\tau) \Lambda(\tau)(\tau) D(\tau), \tau \in {\cal I}\ .
			\end{split}
		\end{equation}
		Here  $\mu(\tau) \in \real^{d}$ is the mean of $x_\alpha(\tau)$ and $e(\tau)$ is an {\it i.i.d.} vector of errors such that $E(e(\tau)) = 0$, $Cov(e(\tau))= I_d$, identity. $D(\tau)$  is a $d \times d$ diagonal matrix denoting conditional standard deviations from univariate GARCH models. $\Lambda(\tau)$ is a time-varying conditional-correlation matrix.
		In DCC-GARCH model, the term $\Lambda(\tau)$ has additional structure that constrains its estimation from the data.  
		Please see Appendix~\ref{app: garch_model} for more information on DCC-GARCH model.
	\end{itemize}

	\subsection{Model Parameter Estimation \& Selection}
\begin{figure}
		\begin{center}
			\includegraphics[height=1.5in, width = 2.5in]{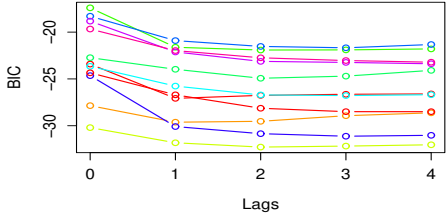}
			\end{center}
			\caption{Model selection: BIC values versus VAR model lags ($p$) for different training sequences.} \label{fig:BIC-curves}
	\end{figure}

\begin{figure}
		\begin{center}
			\includegraphics[height=1.2in, width = 3.5in]{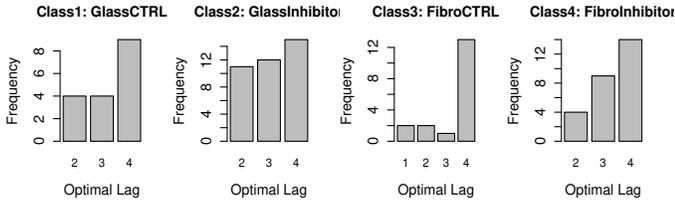}
			\end{center}
			\caption{The four histograms display the frequencies of optimal lags for all shape sequences in our database. Each histogram represents results for a different sequence class (these classes are described later in Section V).} \label{fig:OptimalLag}
	\end{figure}

	Given observed data, the next step is to fit these two models and to evaluate them for the goodness of fit. 
	\begin{itemize}
		\item[I.] {\bf VAR Model}:
		To facilitate estimation of parameters in VAR($p$) model, we rewrite the model as,
		\begin{eqnarray}
			x^{\dagger}_{\alpha}(\tau) = z^{\dagger}(\tau) \boldsymbol{\beta} + \epsilon_{\alpha}^{\dagger}(\tau), \tau \in {\cal I}\ ,
		\end{eqnarray}
		where the superscript ${\dagger}$ denotes transpose, 
		$z(\tau) \doteq (1, x^{\dagger}_{\alpha}(\tau - 1), \cdots, x^{\dagger}_{\alpha}(\tau - p)) \in \real^{1 \times (dp+1)}$ and $\boldsymbol{\beta} \doteq [c; A_1; \cdots; A_p] \in \real^{(dp+1) \times d}$. 
		Let $\boldsymbol{X} = (x_{\alpha}(1), \dots, x_{\alpha}(T))$, $\boldsymbol{Z} = (z(1), \dots, z(T))$, and 
		$\boldsymbol{\epsilon} = (\epsilon_{\alpha}(1), \dots, \epsilon_{\alpha}(T))$.
		Stacking the variables over the observation period and using a vector-matrix notation, we obtain
		$$
		\mbox{vec}(\boldsymbol{X}) = ({I_{T} \otimes Z})\mbox{vec}(\boldsymbol{\beta}) + \mbox{vec}(\boldsymbol{\epsilon})\ ,
		$$
		where $\otimes$ denotes the tensor product and $\mbox{vec}$ is the column-stacking operator.
		Note that the covariance matrix of $	\mbox{vec}(\boldsymbol{\epsilon})$ is $\Sigma_\alpha \otimes {I}_{T}$. 
		The least-square estimate of $\boldsymbol{\beta}$ is obtained by minimizing
		\begin{eqnarray}
			S(\boldsymbol{\beta}) = tr[(\boldsymbol{X} - \boldsymbol{Z \beta}) \Sigma^{-1}_\alpha (\boldsymbol{X} - \boldsymbol{Z \beta})^{\dagger}]\ ,
		\end{eqnarray}
		resulting in: $\hat{\boldsymbol{\beta}} = (\boldsymbol{Z^{\dagger}Z})^{-1}(\boldsymbol{Z^{\dagger}X}) =$
		\begin{equation}
		\left[\sum_{\tau=p+1}^{T} \boldsymbol{z(\tau) z^{\dagger}(\tau)}\right]^{-1} \sum_{\tau=p+1}^{T} \boldsymbol{z(\tau) x^{\dagger}_{\alpha}(\tau)}. \label{eq:var-est}
		\end{equation}
		Note that this estimator does not depend on $\Sigma$. The estimate of covariance matrix $\Sigma_\alpha$ is 
		$$
			\hat{\Sigma}_\alpha = \dfrac{1}{T-(d+1)p-1}(\textbf{X} - \textbf{Z}\hat{\boldsymbol{\boldsymbol{\beta}}})^{\dagger} (\textbf{X} - \textbf{Z}\hat{\boldsymbol{\boldsymbol{\beta}}})\ .
		$$

		The model selection, or the estimation of $p$, is performed using information criteria such as AIC (Akaike information criterion), BIC (Bayesian information criterion), and HQ (Hannan-Quinn information criterion). 
		AIC usually tends to choose large numbers of lags since it asymptotically overestimates the order with positive probability, 
		whereas BIC and HQ estimate the order consistently under fairly general conditions. 
		Figure~\ref{fig:BIC-curves} shows a plot of BIC values for VAR($p$) model versus $p$ for data resulting from 11 shape sequences randomly selected. Figure~\ref{fig:OptimalLag} displays histograms of the optimal lags for 100 shape sequences, separated by their experimental classes. These results show that $p=3$ or $4$ usually provides the best model fit to the data. 
		
		

		\item[II.] {\bf DCC-GARCH Model}: We can estimate The DCC-GARCH model parameters in two steps. The unknown parameter set has two subsets. The first corresponds to the univariate GARCH model parameters for each PCA component, and the second relates to the dynamic correlation part. Note that each element of $x_{\alpha}$ is a scalar time-series of its own. In the first stage, the elements of the first subset are estimated for each series individually. In the second stage, the parameters of dynamic correlation are estimated using an appropriately specified likelihood. Please see Appendix~\ref{app: garch_estimation} for more details on DCC-GARCH model estimation.
		
		Table \ref{tab: garch_estimation} shows an example of some (partial) estimation results using an arbitrary shape sequence. In this example, all the $p$-values of the fitted DCC(1,1)-GARCH parameter are smaller than $0.05$. The long-run persistence for most series is below $0.90$, which indicates it is not a long-memory process. The conditional correlation parameters are significant, implying that the DCC model makes more sense than a {\it constant} conditional-correlation model.
		
	\end{itemize}
	
		\vspace{-2pt}
		
		\vspace{-2pt}

\section{Experimental Evaluation: Shape Synthesis and Prediction}
	In this section, we study the use of our time-series model on TSRVF-PCA representation $x_{\alpha}$ of a shape sequence $\alpha$ in some practical applications -- synthesizing whole new sequences and predicting future shapes in sequences.

	\subsection{Synthesizing Shape Sequences}
	One of the ways of validating a time-series model on shapes is to synthesize new sequences from 
	the model and analyze the results. 
	Let $x_{\alpha}: {\cal I} \to \real^d$ represent TSRVF-PCA representation of an observed shape sequence 
	$\alpha$. Let $\hat{\Theta} = (\hat{\boldsymbol{\beta}}, \hat{\Sigma})$ denote the VAR parameters estimated from the training sequences using generalized least squares, as mentioned above. 
	We can use the model in Eqn. \ref{eqn:var-model}, with the estimated parameters,  to generate new time series	$\tilde{x}_{\alpha}: {\cal I}  \in \real^d$, and then map this synthesized sequence back to the 
	original shape space ${\cal S}$, resulting in a synthesized sequence $\tilde{\alpha}: {\cal I}  \to {\cal S}$. This, of course, requires additional specification of principal scores beyond the first $d$ PC directions and we will set them to be zero to ensure a unique mapping back to ${\cal S}$.
	
\begin{figure}
		\begin{center}
			\begin{tabular}{|c|l|}
				\hline
		Real &	\includegraphics[height=0.2in]{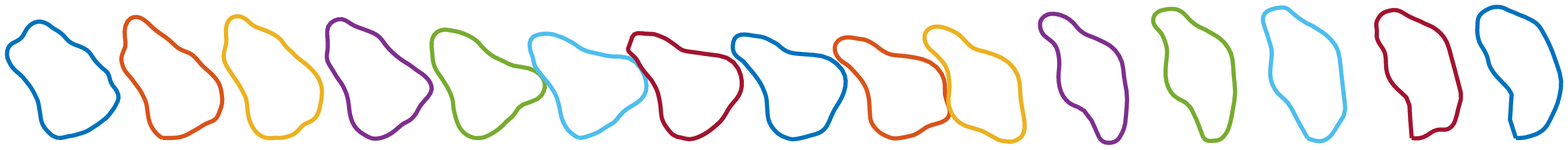} \\
		\hline
		Synthetic 1 &\includegraphics[height=0.2in]{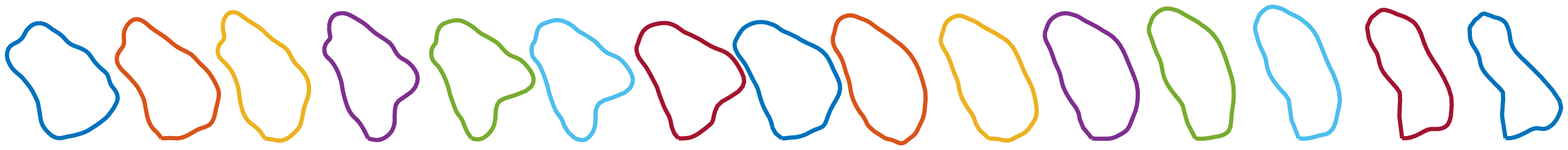} \\
				\hline
		Synthetic 2 &\includegraphics[height=0.2in]{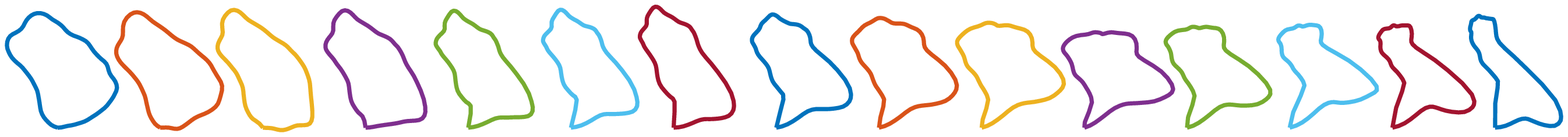} \\
				\hline
		Synthetic 3 &\includegraphics[height=0.2in]{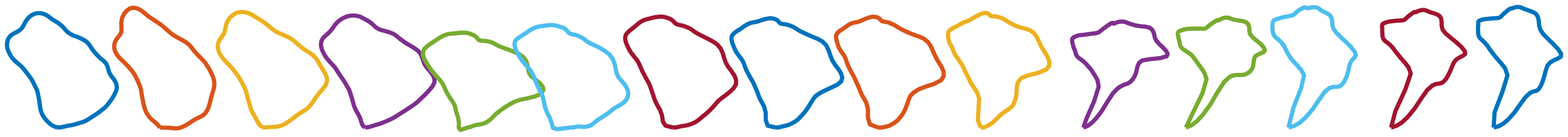} \\
				\hline
			\end{tabular}
			\caption{\small Synthesis of shape sequences using estimated VAR(1) model: 
				Top row shows the original sequence, 
				and the bottom three row show three random synthesized shape sequences, for $d = 5$.} \label{fig:synthesize} 
		\end{center}
	\end{figure}

	We illustrate this idea using some examples in Fig. \ref{fig:synthesize}. The top row shows a part of the the original sequence whose TSRVF-PCA representation is used to fit a VAR(1) model. 
	The bottom three rows show examples of randomly generated shape sequences, for $d=5$, using the estimated model. We initialize each synthesized sequence with the same initial shape $\alpha(0) \in {\cal S}$. The realistic nature of synthesized images provides some authenticity to the 
	proposed dynamical model. This is purely a qualitative evaluation and we will use some quantitative approaches in the next few sections.

	\subsection{Predicting Future Shapes}
	Another way to evaluate a time-series model is to predict future shapes using past sequence data. 
	Consider prediction of $x_{\alpha}(T_0+h)$, for $h > 0$,  using the shapes $\{x_{\alpha}(\tau)| \tau \leq T_0\}$. 
	The best linear predictor of $x_{\alpha}(T_0+h)$ under the $VAR(p)$ model is given by: 
	\begin{eqnarray}
		\hat{x}_{\alpha}(T_0+h|T_0) =  \sum_{j=1}^p \hat{A}_j \hat{x}_{\alpha}(T_0+h-j|T_0)
	\end{eqnarray}
	where $\{\hat{A}_j\}$ are the estimated matrices. Note that the $h$-step prediction 
	involves predicting all the intermediate shapes from $T_0+1$  to $T_0 + h$. If the ground truth shapes at $T_0+1, \dots, T_0+h$ are known, then we can quantify the prediction error using: 
	${ 1 \over h} \sum_{i=1}^h \| x_{\alpha}(T_0+i) - \hat{x}_{\alpha}(T_0+i|T_0) \|^2$.
	
\begin{figure}
		\centering
		\begin{tabular}{|c|c|}
			\hline
			Lag $p$ & Forecasting shape sequences \\
			\hline
			1  & \includegraphics[height=0.22in]{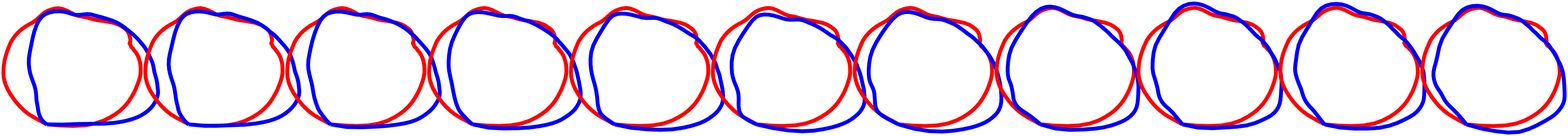} \\
			\hline
			2 & \includegraphics[height=0.22in]{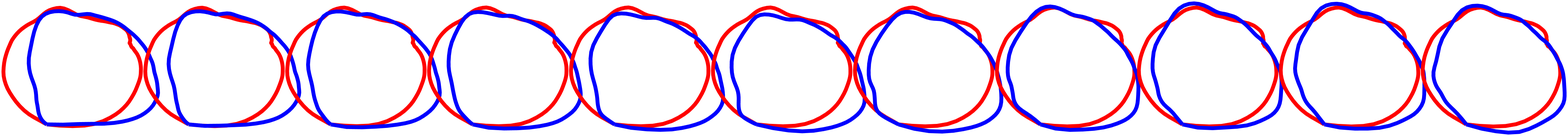} \\
			\hline
			3 & \includegraphics[height=0.22in]{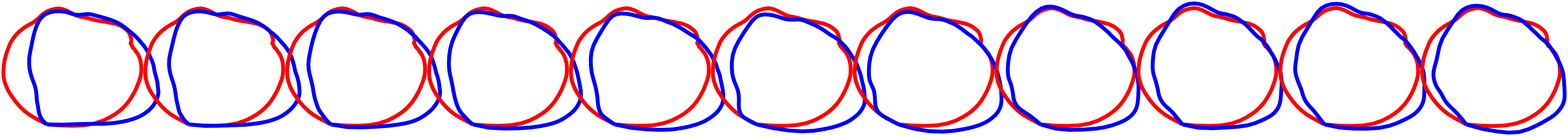} \\
			\hline
			4 & \includegraphics[height=0.22in]{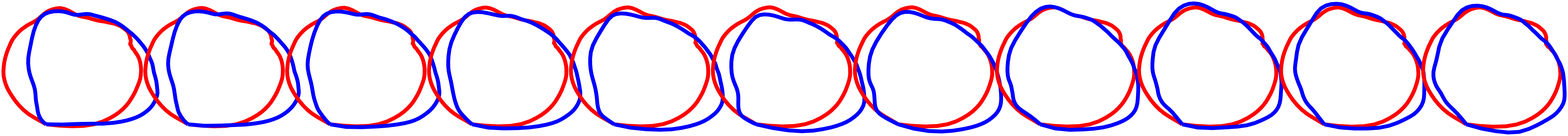}\\
			\hline
		\end{tabular}
		\begin{tabular}{cc|c|c|}
			\includegraphics[height=1.3in]{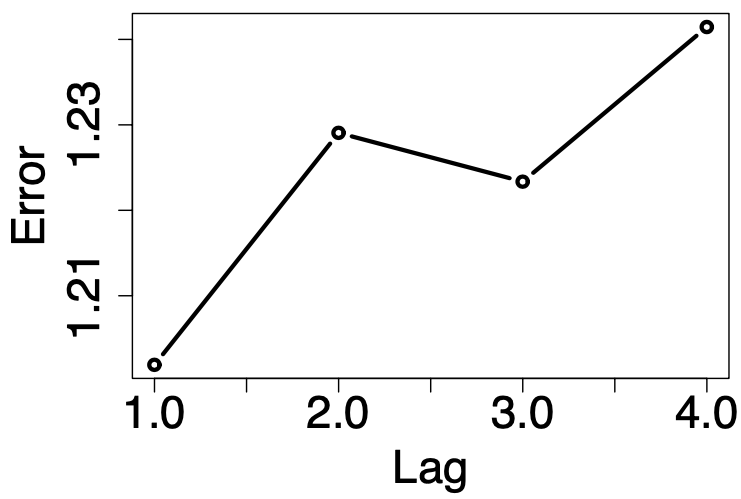}
		\end{tabular}
		\caption{Top: Prediction and associated errors of a shape sequence $\alpha$ from its Euclidean representations in $\real^d$ for different lags. The solid contours are real shapes and the dashed contours are predicted shapes. Bottom: Prediction error versus model lag $p$.} 
		\label{fig:forecastingComparison}	
	\end{figure}

	We present some experimental results on predicting future shapes using the estimated VAR model. We use an arbitrarily selected sequence to demonstrate this idea. Figure ~\ref{fig:forecastingComparison} shows the prediction results for the specified sequence for different values of the lag parameter $p$. We use $d = 5$ to form a TSRVF-PCA representation of the shape sequence in each case.  
	In this example, the original sequence has $179$ shapes with time points ${\cal I} = \{1,2,...,179\}$. So the resulting Euclidean time-series $x_{\alpha}$ has the size $5 \times 178$. We split the sequence into a training set $(\tau = 1,2,...,134)$ and testing set $(\tau = 135, ..., 178)$, estimate parameters for VAR($p$) model using the training set,  and then predict the test sequence recursively. 
	
	Figure~\ref{fig:forecastingComparison} shows some results from this experiment. 
	The top four rows show examples of predicted shapes overlaid on the true shape for different values of $p$. For each sequence, the red curves denote the predicted shapes overlaid on the true shapes. On close inspection, one can see that the prediction error for VAR($1$) is smaller than that for other lag values. The bottom plot quantifies this and shows prediction errors for different values of $p$ on the test part of this sequence. 
	The results indicate that $p=1$ provides the best performance. Although these results were generated using one shape sequence, they are typical of results obtained from several sequences. 
	
 	\subsection{Model Selection}
	Since we have multiple models (VAR and DCC-GARCH), we need to select a model for the subsequent analysis using shape data.  
	For shape sequences in our database, we first estimate model parameters for different models from the shape data and then evaluate these competing models in terms of prediction errors $E_\alpha = \left\| \hat{x}_\alpha - x_\alpha \right\|$.
	
\begin{figure} 
		\centering
		\begin{tabular}{|c|c|}
			\hline
			Model & Forecasting shape sequences\\
			\hline
			VAR(1) & \includegraphics[height=0.22in]{figs/k39_var_fcst_lag2.eps} \\
			\hline
			DCC(1,1)-GARCH &  \includegraphics[height=0.22in]{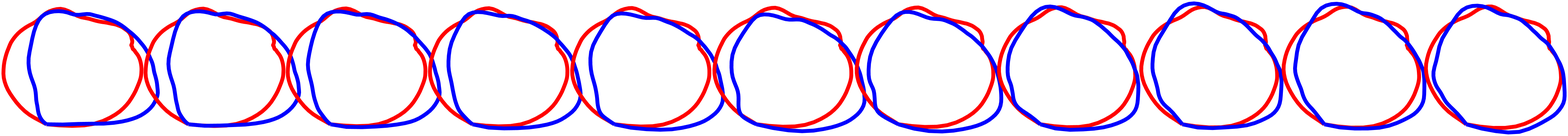} \\
			\hline
		\end{tabular}
		\caption{This table compares forecasting errors for VAR and GARCH models. The red curves denote the predicted shapes and blue curves denote the ground truth shapes.} 
		\label{fig: model_comparison}
	\end{figure}

	In Fig.~\ref{fig: model_comparison}, we display an example of predicting future shapes from the past data under VAR(1) and DCC(1,1)-GARCH model. Table \ref{tab: model_comparison} quantifies the prediction errors for different shape sequences. It shows that the errors for the VAR model are smaller than those for the GARCH model. Additionally, the VAR model is relatively simpler and computationally more efficient than the DCC-GARCH model. Thus, we will restrict to a VAR model in all subsequent analyses. 

	\begin{table}[!htbp]
		\centering
		\caption{Prediction errors of shape sequences $\alpha$ from its Euclidean representations in $\real^d$ for VAR model and GARCH model.} 
		\label{tab: model_comparison}
		\begin{tabular}{|c|c||c|c|c|c|c|}
			\hline
			\multicolumn{2}{|c||}{Sequence} & 1 & 2 & 3 & 4 & 5 \\
			\hline
			\multirow{2}{*}{Errors} & VAR  & 1.299 & 1.247 & 2.757 & 2.276 & 1.307 \\ 
			\cline{2-7} 
			& GARCH  & 1.622 & 1.940 & 2.846 & 2.375 & 1.712 \\ 
			\hline
		\end{tabular}
	\end{table}


	
	\section{Experimental Evaluation: Classification of shape dynamics}
	Using the pipeline laid out in Fig.~\ref{fig:pipeline}, we have developed an efficient Euclidean representation for shape sequences and imposed time series models on them.
	So far, we have utilized time-series models in generative tasks -- shape synthesis and shape prediction. Another important use of a statistical model is in classification, and we study that task next. 
	
	\subsection{Problem Challenges and Overview}
	The dynamical shapes of cells in different biological experimental protocols are governed by underlying conditions that dictate specific behavior. However, unlike prevalent image classification problems, where it is possible to visualize clear distinctions across classes manually, the dynamics of cellular shapes have no easily perceivable feature. To highlight differences in the cell motility under different experimental conditions -- in the presence or absence of ROCK inhibitor, in the glass or fibronectin medium -- we utilize our $VAR(p)$ and perform statistical classification. 
	
	We demonstrate the efficacy of the developed method using both simulated sequences and real-world data. 
	To classify shape evolution into predetermined classes of cell migration, each representing a different experimental setting, we extract various features representing shape dynamics and then plug these features in standard classifiers to perform classification. 
	
	\subsection{Features for Classification}
	For our experiments and evaluation, we will use both shape and kinematic features, individually and jointly, to perform sequence classification. \\
	
    \noindent {\bf Dynamic shape features}:
    Representation of shape sequences by Euclidean time-series (TSRVF-PCA) enables us to explore several candidate features for classification. For instance, one can concatenate elements of this times series $\{x_{\alpha}(\tau), \tau \in {\cal I}\}$ into a long vector. Another possibility is to use parameters $\Theta$ of the VAR model, fitted to the time series $x_{\alpha}$, as features. Although this representation is indirect, using model parameters instead of observations, it provides a more compact representation. A direct comparison of $x_{\alpha_i}$ and $x_{\alpha_j}$ using {\it e.g.,} a Euclidean metric is prone to temporal misalignment issues. The two sequences can come from the same class but maybe at different dynamic stages at $\tau = 1$. In contrast, the VAR parameters capture {\it modes} of shape dynamics and are invariant to such nuisance variables as the temporal registration and sequence lengths. 
    
    Choosing VAR parameters of a sequence as its features, we represent each shape trajectory $\alpha$ by $\Theta_{\alpha}$ estimated from the intermediate Euclidean representation $x_{\alpha}$.  Any two shape sequences are then compared using a Euclidean distance ${d}_{\alpha}(i, j) = \| \Theta_{\alpha_i} - \Theta_{\alpha_j}\|$. When VAR model parameters are estimated using same lag value for each sequence, \textit{i.e.}, $p_i = p_j$, $d_{\alpha}$ is computed using Eqn.~\eqref{con:distance1}. However, different values of lag may be optimal for different sequences (as observed in Fig.~\ref{fig:OptimalLag}). Thus, we need a way to compare the VAR model parameters associated with different lags across sequences. In that case, we use Eqn.~\eqref{con:distance2} to compute $d_{\alpha}$. In addition, we also explore features that combine all lag values. A feature in this case is generated by concatenating $\Theta_{\alpha_i}$ obtained for $p_i = 1, \dots, 5$. The distance vector $d_{\alpha}$ is then computed using Eqn.~\eqref{con:distance1} using the concatenated VAR model parameters. This process results in the final shape feature, denoted by $\mathcal{V}_S \in \real^{N_0}$, by concatenating these distance vectors, as described in step:7 of Algorithm 1.
	 \\
	 	 \noindent {\bf Kinematics features}: Cell kinematics (such as velocity, angular velocity and others) can also provide valuable information in categorizing modes of cell motility. To demonstrate and compare the efficiency of kinematics {\it versus} shape features in classification, we also derive some kinematics features from contours.  
	 	 These features include: vector of instantaneous speeds $\{\upsilon(\tau) \in \real^2\}$, vector of instantaneous scales $\{\eta(\tau) \in \real \}$, a vector of five-step rotations $\xi \in \real^5$. Here $\upsilon(\tau)$ is defined as the change in the location of the center of mass and $\eta(\tau)$ is the change in the perimeter of the contour from time $\tau$ to time $\tau+1$. 
	 Let $\theta_h(\tau)$ be the rotational alignment between contours at time $\tau$ and $\tau+h$, $h = 1,...,5$, then we set $\xi_{h} = \dfrac{1}{T-h}\sum_{\tau=1}^{T-h}\theta_h(\tau)$. 
	 The final kinematics feature of the sequence $\alpha$ is obtained as,
	 $\Omega_{\alpha} = [\upsilon, \eta, \xi] \in \real^{3T + 5}$. 
	 We compare any two contour sequences using a {\it kinematics-based distance}: $ d_{\Omega}(i, j) = \left\| \Omega_{\alpha_i} - \Omega_{\alpha_j}  \right\|$, where $\Omega_{\alpha}$ are the kinematics features of a contour sequence $\alpha$. 
	 The feature distance between any two sequences $\{ {d}_{\Omega}(i, j)\}$ is then computed using Eq.~\ref{con:kine_distance}. The final kinematics features $\mathcal{V}_K \in \real^{N_0}$ is obtained by concatenating the distances between a sequence and all training sequences, as described in step:7 of Algorithm 1.

	Thus, each sequence $\alpha$ is represented by a distance vector $\mathcal{V}_S$ or $\mathcal{V}_K  \in \real^{N_0}$ where
	$N_0 = Cm$, $C$ is the number of classes and $m$ is the number of training sequences in each class.
	These distance-feature vectors are 
	used in conjunction with different classifiers to determine the classification performance. Algorithm \ref{alg:Classification} provides a step-by-step layout of the complete classification procedure.
	
	\subsection{Classifiers}
	Classification is performed using several current classifiers: Support Vector Machine (SVM), Random Forest, Decision Tree, and Convolutional Neural Network (CNN). For the CNN classifier, we employ a simple $1$D Convolutional Neural Network to classify $1$D  feature vectors. 
	The neural network architecture consists of four convolutional layers with kernel size $3$. Each convolution is sequentially followed by Batch-Norm, ReLU \cite{hara2015analysis} activation, dropout and $1$D pooling layer.
	The network is trained using the cross-entropy loss for classification, and the network weights are updated using the Adam  \cite{kingma2014adam} optimization. It is trained for $15$ epochs with a batch size of $10$ and a learning rate of $0.01$. 
	
	\begin{algorithm}[htb] 
		\caption{Classification} 
		\label{alg:Classification} 
		\begin{algorithmic}[1] 
			\REQUIRE 
			Euclidean time series $x_{\alpha_i} \in (\real^d)^{\cal I}$ for shape sequences $\alpha_i$, $i = 1,...,N$.\\
			\STATE Split data  into training set of size $N_0 < N$ and testing set of size $N- N_0$.
			\STATE Estimate VAR$(p)$ model for each sequence $x_{\alpha_i}$, and set parameter $\Theta_{\alpha_i} = (c, A_1, ..., A_p, \Sigma_\alpha)$.
			\STATE Let $\boldsymbol{\beta}_\alpha = (c,A_1, ..., A_p)^{\dagger} \in \real^{(dp+1) \times d}$, and $\bar{A}_\alpha = \dfrac{1}{p}\sum_{i=1}^{p} A_i \in \real^{d \times d}$. \\
			\STATE Compute kinematics feature $\Omega_i$ for each sequence $\alpha_i$. Make these features vectors to be of the same size using zero padding. That is, if $T_{\alpha_i} > T_{\alpha_j}$, set $\upsilon_{\alpha_j}  = (\upsilon^{\dagger}_{\alpha_j}, \textbf{0})^{\dagger}.$ 
			\STATE Normalize these features using Euclidean norms according to: \\
			$\boldsymbol{\beta}^*_\alpha = \dfrac{\boldsymbol{\beta}_\alpha}{\left\| \boldsymbol{\beta}_\alpha\right\| }$, 
			$\bar{A}^*_\alpha = \dfrac{\bar{A}_\alpha}{\left\| \bar{A}_\alpha\right\| }$, 
			$\Sigma^*_\alpha = \dfrac{\Sigma_\alpha}{\left\| \Sigma_\alpha\right\| }$; \\
			$\xi^*_\alpha = \dfrac{\xi_\alpha}{\left\| \xi_\alpha\right\|} $,
			$\upsilon^*_\alpha = \dfrac{\upsilon_\alpha}{\left\| \upsilon_\alpha\right\|} $,
			$\eta^*_\alpha = \dfrac{\eta_\alpha}{\left\| \eta_\alpha\right\|} $.
			\STATE Compute pairwise distance between training sequences:
			$ d_{\Omega}(i, j) = \left\| \Omega_{\alpha_i} - \Omega_{\alpha_j}  \right\| $
			\begin{numcases}{=}
			\sqrt {\left\| \xi^*_{\alpha_i} - \xi^*_{\alpha_j} \right\|^2 + \left\| \upsilon^*_{\alpha_i} - \upsilon^*_{\alpha_j} \right\|^2 + \left\| \eta^*_{\alpha_i} - \eta^*_{\alpha_j} \right\|^2} \ .	
			 \label{con:kine_distance}			
			\end{numcases}			
			$	d_{\alpha}(i, j) = \left\| \Theta_{\alpha_i} - \Theta_{\alpha_j} \right\| $				
			\begin{numcases}{=}
				\sqrt{\left\| \boldsymbol{\beta}^*_{\alpha_i} - \boldsymbol{\beta}^*_{\alpha_j} \right\|^2 + \left\| \Sigma^*_{\alpha_i} - \Sigma^*_{\alpha_j} \right\|^2 }, &$p_i  = p_j$ \label{con:distance1}\\
				\sqrt{\left\| \bar{A}^*_{\alpha_i} - \bar{A}^*_{\alpha_j} \right\|^2 +\left\| \Sigma^*_{\alpha_i} - \Sigma^*_{\alpha_j} \right\|^2}, &$p_i  \neq p_j$ .\label{con:distance2}
			\end{numcases}						
			\STATE For each training sequence $x_{\alpha_i}$, the final shape feature vector is obtained as $\mathcal{V}_{S_i} = [d_\alpha(i,1), \dots d_\alpha(i,N_0)]$  and kinematics features as $\mathcal{V}_{K_i} = [d_\Omega(i,1), ..., d_\Omega(i,N_0)]$. Here $N_0$ is the total number of training sequences.
			\STATE $\mathcal{V}_S$, $\mathcal{V}_K$ and $\mathcal{V}_{S,K} = w_1 \mathcal{V}_S + w_2 \mathcal{V}_K$ are used to train the classifier.
			\STATE In the testing phase, $\mathcal{V}_S$, $\mathcal{V}_K$ and $\mathcal{V}_{S,K}$  are computed in the similar manner for a sequence in the test dataset and evaluated using respective trained classifier models.
	
		\end{algorithmic}
	\end{algorithm}

	

	\subsection{Classification Performance on Simulation Data}
	To evaluate classification performance using the proposed framework, we first perform experiments on a simulated dataset.
	
	\subsubsection{Simulation Procedure}: We simulate shape sequences for four classes that correspond to four combinations of experimental conditions:  using glass/fibronectin support and with/without ROCK inhibitor.

\begin{figure}
		\begin{center}
			\begin{tikzpicture}
				\node (a) {$y$};
				\node (b) [below of = a, yshift = 1.75cm, right of = a, xshift = 0.1cm ] {$y_1$};
				\node (c) [below of = b, yshift = -.15cm] {$y_2$};
				\node (d) [right of = b, xshift = 0.6cm] {Coeff. $c_1$};
				\node (e) [right of = c, xshift = 0.6cm] {Coeff. $c_2$};
				\node (f) [right of = d, xshift = 1.3cm] {VAR Model};
				\node (j) [right of = f, xshift = 0.7cm] {$\tilde{c}_1$};
				\node (l) [right of = j, xshift = 0.1cm] {$\tilde{y}_1$};
				\node (g) [right of = e, xshift = 1.3cm] {VAR Model};
				\node (k) [right of = g, xshift = 0.7cm] {$\tilde{c}_2$};
				\node (m) [right of = k, xshift = 0.1cm] {$\tilde{y}_2$};
				\node(i) [below of = l, yshift = .4cm, right of = l, xshift = 0.1cm] {$\tilde{y}$};		
				\draw [arrow](a) -- (b);
				\draw [arrow](a) -- (c);
				\draw [arrow](b) -- (d);
				\draw [arrow](c) -- (e);	
				\draw [arrow](d) -- (f);
				\draw [arrow](e) -- (g);
				\draw [arrow](f) -- (j);
				\draw [arrow](g) -- (k);
				\draw [arrow](k) -- (m);
				\draw [arrow](m) -- (i);
				\draw [arrow](l) -- (i);
				\draw [arrow](j) -- (l);
			\end{tikzpicture}
		\end{center}
		\caption{Simulation steps based on imposing VAR models on Fourier coefficients of the coordinate functions of contours.} 
		\label{fig:simulation}
	\end{figure}
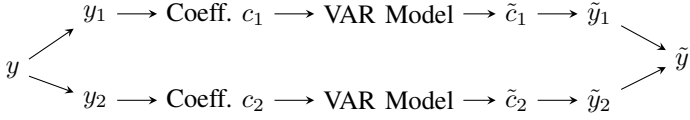

    The simulated sequences are generated as follows. 
	We randomly select one real observed contour sequence from each class.
	As shown in Fig.~\ref{fig:simulation}, we take a 2D contour $y(\tau)$ at time $\tau$, and consider its coordinates functions denoted by $y_1(\tau)(t)$ and $y_2(\tau)(t)$, as separate scalar functions. We use the Fourier basis ${\cal B} = \left\lbrace 1, \sqrt{2}\sin(2\pi nt), \sqrt{2}cos(2\pi nt), n=1,2, \dots,m \right\rbrace$ to express these functions via their coefficients: $y_1(\tau)(t) = \sum_{b \in {\cal B}} c_1(\tau,b) b(t)$ and 
	$y_2(\tau)(t) = \sum_{b \in {\cal B}} c_2(\tau,b) b(t)$. 
	Then, we fit an $m$-dimensional VAR model, where $m$ is the number of basis elements used, to these coefficients $c_1$ and $c_2$ separately and estimate the corresponding VAR parameters (note that these model estimations are different from the VAR fittings performed for TSRVF-PCA coefficients discussed earlier). 
	Using these estimated parameters, we generate new, random time-series of Fourier coefficients $\{\tilde{c}_1(\tau), \tilde{c}_2(\tau), \tau =1,2,\dots, T\}$ and reconstruct full contours using 
	$\tilde{y}_1(\tau) = \sum_{b \in {\cal B}} \tilde{c}_1(\tau,b) b$ and 
	$\tilde{y}_2(\tau) = \sum_{b \in {\cal B}} \tilde{c}_2(\tau,b) b$. A schematic for the simulation procedure is shown in Fig.~\ref{fig:simulation}.

\begin{figure}
		\begin{center}
			\begin{tabular}{|c|}
				\hline
				\includegraphics[width=3.4in]{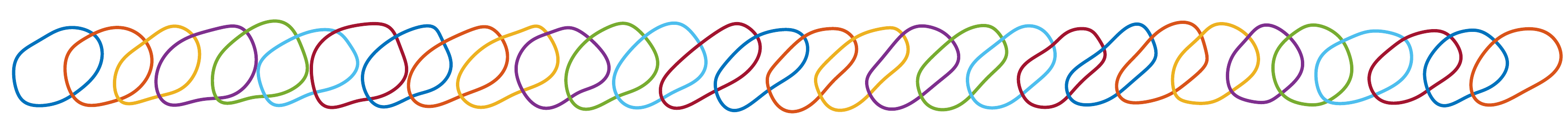} \\
				\includegraphics[width=3.4in]{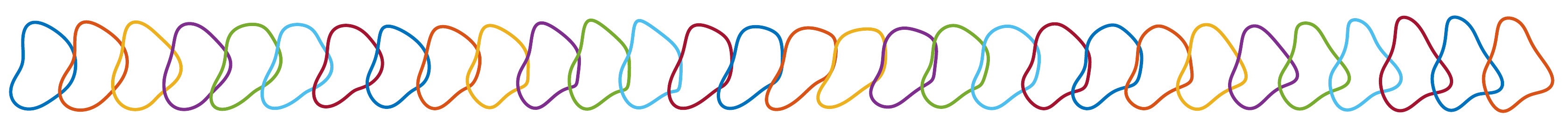} \\
				\includegraphics[width=3.4in]{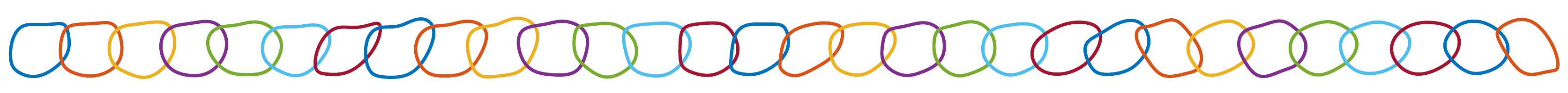} \\
				\includegraphics[width=3.4in]{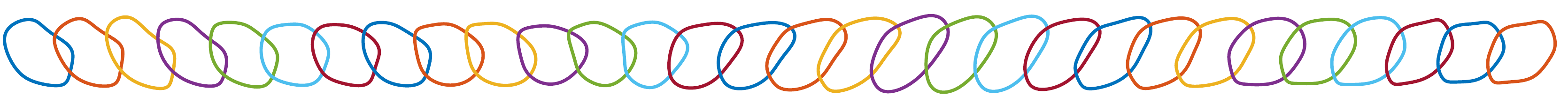} \\
				\hline
			\end{tabular}
			\vspace{-3pt}
			\caption{Examples of simulated contour sequences, one for each class, 
				simulated using parameters estimated from real data.} \label{fig:simdata}
		\end{center}
	\end{figure}

	\subsubsection{Classification using simulated data} Using the simulation mentioned above, we generate 400 simulated contour sequences for four classes -- 100 sequences for each class -- with each sequence being of length $T = 150$. Figure~\ref{fig:simdata} shows examples of these simulated sequences, one for each class. Given this simulated data, we perform a five-fold classification experiment as follows. In every run, we use 80\% of the sequences per class for training and 20\% for testing. We use their TSRVF representations to perform PCA in the training phase, resulting in a Euclidean time series for each shape sequence (in the TSRVF-PCA space with $d = 5$). We project the TSRVFs of the test sequences in the same PCA space. Next, we estimate the VAR parameters for each Euclidean time series -- $4 \times 80$ training and $4 \times 20$ test -- separately. As mentioned earlier, we use a distance between these VAR parameters to compare and classify sequences. We study different potential values of the VAR lag to optimize the classification rate. For any sequence, $x_\alpha$, either test or training -- we use its distance vector $\mathcal{V}_{\alpha}$ computed from each of the $80$ training sequences and arrange in a vector of size $320$. This set of distances denotes the feature vector for classification. 
	
	\begin{table}[!ht]
			\centering
			\caption{The average classification rates for simulated data.} 
			\begin{tabular}{|c||c|c|c|c|c|c|c|c||}		
				\hline
				Classifier &\multicolumn{5}{c|}{ SVM}   &CNN\\
				\hline
				VAR Model Lag & 1 & 2 & 3 & 4 & 5  &  1-5\\
				\hline
					{\tabincell{c}{Class. Rate (\%)}}
					& \textbf{0.940 }& 0.865 & 0.795 & 0.760  & 0.760 &0.920 \\
				\hline		
			\end{tabular}
			\label{tab:class_sim}
		\end{table}
	
	Table~\ref{tab:class_sim} presents the average classification rates on this simulated data for different values of the 
	VAR lag parameter $p$. The VAR(1) model yields the highest accuracy of 94\% in this four-class setup using the SVM classifier, demonstrating that the designed shape-dynamics model can be effective in classifying dynamic shape sequences.

	\subsection{Classification Evaluation on Microscopy Data}
	Next, we apply this classification method to the microscopy data and study various modes of motility for  \textit{E. histolytica}. We hypothesize that a change in  experimental conditions causes a change in the motility patterns and investigate that hypothesis using real data.  
	
	\begin{figure}
		\centering
		\includegraphics[width=0.8\linewidth]{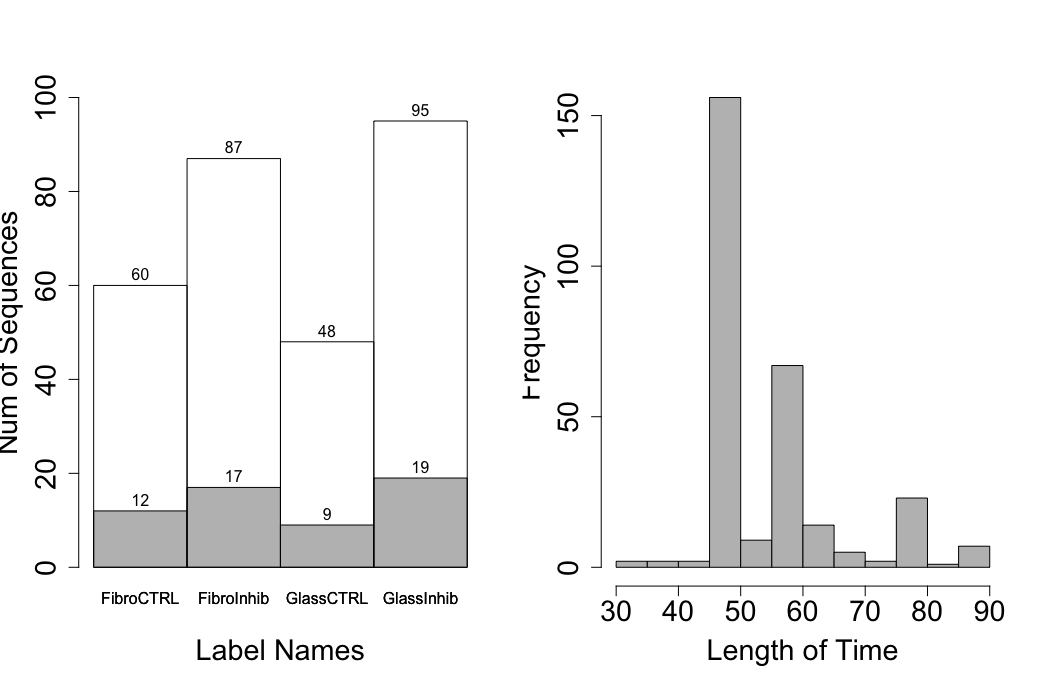}	
		\caption{Histograms displays microscopy data information. The left figure shows the number of sequences for each class. The grey bins show the counts for testing set and the white bins show the counts for original data set. The right figure shows the length of time for all TSRVF-PCA sequences}
		\label{fig:DataInfo}
	\end{figure}

	The computation of TSRVF-PCA vectors, VAR parameter estimation, and representation of sequences as distance vectors between VAR parameters remain the same as described in the simulation study. The computation is performed on the cell contours extracted from the videos, as described earlier in the paper.
	The features are extracted for $290$ sequences in total, divided randomly into a training set ($233$ sequences, consisting of $80\%$ data from each class) and a testing set ($57$ sequences, approximately $20\%$ per class). The total numbers of sequences for each class are shown in Fig \ref{fig:DataInfo}. 
	

	\begin{figure}
		\centering
		\includegraphics[width=0.5\linewidth]{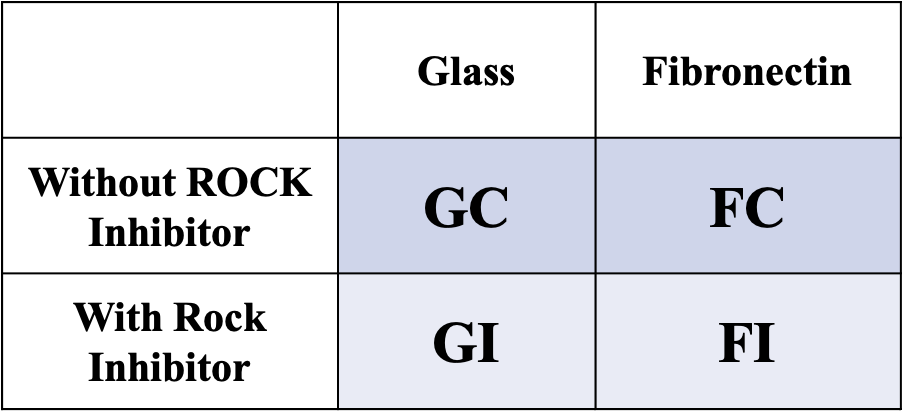}	
		\caption{Class label information for microscopy data analysis}
		\label{fig:classInfo}
	\end{figure}
	\vspace{-2pt}
	
    The categories for classification are defined for different biological experimental conditions as shown in Fig.~\ref{fig:classInfo} (e.g., the matrix block 'GC' denote the category for shapes extracted from cells on glass without inhibitor). The classification tasks are defined based on a different grouping of these biological experiments. First, we perform a multi-class classification with four classes indicating motility of \textit{Entamoeba} on glass and fibronectin with and without the ROCK inhibitor {\it i.e.}, with categories 'GC', 'GI', 'FC', and 'FI' (Fig.~\ref{fig:classInfo}). In the second experiment, we perform classification between broader categories: amoeba on glass or fibronectin {\it i.e.}, 'GC' and 'GI' versus 'FC' and 'FI' (grouping data along columns in Fig.~\ref{fig:classInfo}). Similarly, for the third experimental classification of amoeboid motion with or without inhibitor is performed, i.e., 'GC' and 'FC' versus 'GI' and 'FI' (grouping data along columns in Fig.~\ref{fig:classInfo}). Table.~\ref{tab:class_real} presents results of classification in these three experiments. 
    
    The final set of experiments involves pairwise classification between finer class labels: (i) amoeba on glass with and without inhibitor ('GC' vs. 'GI'), (ii) on glass vs. fibronectin, both without inhibitor ('GC' vs. 'FC'), (iii) on glass vs. fibronectin with inhibitor ('GI' vs. 'FI'), and (iv) fibronectin without or with inhibitor ('FC' vs. 'FI'). The classification results of these four experiments are presented in Table.~\ref{tab:class_real_two_class}.
	
	Selecting the VAR model parameter is a crucial design choice in computing the dynamic shape features. As demonstrated earlier (in Fig.~\ref{fig:OptimalLag}), the optimal VAR  model lag for the shape sequence representation varies between sequences. We perform classification experiments with features computed in multiple ways: using different lag parameters (ranging from 1 $\dots$ 5), using the optimal (Best) lag for each sequence, and using all lag values combined. The classification is performed using the dynamic shape features ($\mathcal{V_S}$) either individually or in combination with kinematics ($\mathcal{V}_K$) features  (comb.). The accuracy for the classification tasks is presented in Table.~\ref{tab:class_svmparameterfourclass}. It is observed that almost in all the cases, either the features computed using an optimal lag parameter or that obtained by combining all the lags provides the best performance. Based on this observation, the final table for classification accuracy using $\mathcal{V}_K$ and $\mathcal{V}_S$, individually as well as combined, are presented in Table.~\ref{tab:class_real} and Table.~\ref{tab:class_real_two_class}.
	\\

	\noindent {\bf Remarks on Multi-Class Classification}:
    In this experiment, we consider all four biological classes denoted as: 'GC', 'GI", 'FC', 'FI' (refer Fig.~\ref{fig:classInfo}). The classification accuracy using $\mathcal{V}_K$, $\mathcal{V}_S$ and $\mathcal{V}_K + \mathcal{V}_S$ with different classifiers is presented in Table.~\ref{tab:class_real} (A. Multi Class).
	The neural network yields the best overall accuracy of $86.3\%$ with combined lag representation while using both shape and kinematics features. We note that the proposed method to capture the dynamics of evolving shapes (TSRVF-PCA-VAR) performs significantly better when compared to the TSRVF-PCA features. This improvement underlines the fact that the proposed shape features indeed capture the shape dynamics more efficiently. Table~\ref{tab:confisionmatrix2class} presents the confusion matrices for both the neural network and the SVM classifier. 
	
    We make certain observations from Table~\ref{tab:confisionmatrix2class}. When using shape features, around 11\% of glass control ('GC') amoeba were recognized as glass inhibitors (GI). The number dropped to 7\% with the combined shape and kinematic features. In both scenarios, none were classified as fibronectin sequences. For amoeba migrating on fibronectin (FC), 10\% (with $\mathcal{V}_S$) and only 3\% ($\mathcal{V}_K + \mathcal{V}_S$) are classified as fibronectin inhibitor (FI). Similar to the first observation, none of the amoebae on fibronectin were confused with that on glass. In contrast, the amoeba population inhibited for ROCK activity and moving on glass (GI) are mis-classified as fibronectin (FC) by 16\% (or 17\%) when using shape (or combined) features. Likewise, amoeba on fibronectin and inhibited for ROCK activity (FI)  are mis-classified by 21\% (or 14\%)  to the population on glass (GC) with shape-only (or combined) features.
	\\

	\noindent {\bf Remarks on Binary Classification}:
	The classification accuracy for the broader binary classification is performed in two ways: (i) between amoeba migrating on glass and fibronectin (data is grouped along the columns in Fig.~\ref{fig:classInfo}) (ii) between amoeba with and without ROCK inhibitor (data is grouped along the rows in Fig.~\ref{fig:classInfo}). 
	The mean classification accuracy for these two experiments are presented in Table~\ref{tab:class_real} column B and C respectively using kinematics and shape features separately and combined. We obtain 94.8\% and 87.2\% accuracy using the neural network for the two experiments, respectively. The classification accuracy is significantly higher when using either the shape features or shape and kinematics combined compared to just kinematics features. This implies that the shape evolution pattern is more distinct between amoebas migrating on glass and fibronectin.
	
	The final binary classification experiments were performed on finer class labels to determine any commonality in the migration patterns (when looked into the experimental conditions individually). The classification accuracy for these pair of experiments are presented in Table~\ref{tab:class_real_two_class}. We achieve significantly high classification accuracy for the chosen classification tasks using the combined VAR model lag setting. While using a combination of shape and kinematics features achieves better classification accuracy, the designed shape features just by themselves achieve comparable performance.

\begin{table*}[h!]
	\centering
	\caption{The average classification rates for microscopy data.} 
	\label{tab:class_real}
	\begin{tabular}{|c|c||c|c|c||c|c|c||c|c|c||}	
		\hline
		\multirow{3}{*}{Classifier}	& \multirow{3}{*}{\tabincell{c} {VAR \\ Model \\ Lag}} & \multicolumn{9}{c||}{Class. Rate (\%)}   \\ \cline{3-11}
		& & \multicolumn{3}{c||}{\textbf{A. }Multi Class} &  \multicolumn{3}{c||}{\tabincell{c}{\textbf{B.} Two Class: \\ Glass vs. Fibronectin}} & \multicolumn{3}{c||}{\tabincell{c}{\textbf{C.} Two Class: \\ Inhibitor vs. w/o Inhibitor}}  \\	\cline{3-11}
		&   &  Kinematics & Shape & Combined & Kinematics & Shape & Combined & Kinematics & Shape & Combined     \\ \hline
		\tabincell{c}{TSRVF-PCA \\+ Conv. NN} & -- & -- & 0.341 & --& -- &0.513 & -- & -- & 0.632 &--   \\ \hline
		\tabincell{c}{TSRVF-PCA-VAR \\+ SVM} & -- &  0.368 &0.365  & 0.512& 0.649  & 0.512 & 0.660  &  0.628 & 0.649 & 0.719 \\ \hline
		\tabincell{c}{TSRVF-PCA-VAR \\+Random Forest} 	& Best Lag  &  0.372 &0.337  & 0.537 & 0.604  & 0.544 & 0.691  &  0.646 & 0.625 & 0.717 \\ \hline
		\tabincell{c}{TSRVF-PCA-VAR \\ Decision Tree} &  Best Lag &   0. 333 &  0.319  & 0.425 &  0.544 & 0.533 &0.642  &  0.600 & 0.565 & 0.653  \\ \hline
		\tabincell{c}{TSRVF-PCA-VAR \\+ Conv. NN} & Best Lag & \multirow{3}{*}{0.769} & 0.835 & 0.849 & \multirow{3}{*}{0.879} & 0.904&\textbf{0.948}  & \multirow{3}{*}{0.753} & 0.760 & 0.841  \\ \cline{1-2} \cline{4-5} \cline{7-8} \cline{10-11} 
		\tabincell{c}{TSRVF-PCA-VAR \\+ Conv. NN} & \tabincell{c}{All Lag\\(1-5)} & & 0.811& \textbf{0.863} & & 0.902 & 0.920 &  & 0.865 & \textbf{0.872}    \\ \hline
		
	\end{tabular}
\end{table*}

\begin{table*}[!htb]
	\centering
	\caption{Classification accuracy for different lag parameters on microscopy data using TSRVF-PCA-VAR features (shape) and TSRVF-PCA-VAR  + kinematics features (Comb.)  with SVM for different experimental setup comparison.} 
 	\label{tab:class_svmparameterfourclass}
	\begin{adjustbox}{width=\linewidth}
	\begin{tabular}{|c||c|c||c|c||c|c||c|c||c|c||c|c||c|c|}	
		\hline
		  
		  \multirow{2}{*}{\tabincell{c} {VAR \\ Model Lag}} & \multicolumn{2}{c||}{GC vs. GI } & \multicolumn{2}{c||}{GC vs. FC } & \multicolumn{2}{c||}{ GI vs. FI} & \multicolumn{2}{c||}{ FC  vs. FI} &\multicolumn{2}{c||}{ Glass vs. Fibro} & \multicolumn{2}{c||}{Inhib. vs. w/o Inhib.} & \multicolumn{2}{c|}{Multi Class} \\	\cline{2-15}

 		  & Shape & Comb. & Shape &  Comb.& Shape &  Comb. & Shape &  Comb. & Shape & Comb. & Shape &  Comb.& Shape &  Comb.   \\ \hline
 		  1 & 0.579 & \textit{\textbf{0.793}}  & 0.486 &0.686  & 0.506  & 0.572 & 0.607 & 0.669 & 0.488 &  0.656 & 0.547 & 0.723  & 0.340 & 0.467 \\ \hline 
 		 2 	&  0.657 & \textit{\textbf{0.793}}    & 0.486 & 0.667     & 0.478 & \textit{\textbf{0.661}}   &  0.531 & 0.669	   	 & 0.463 & 0.653  & 0.642 & 0.712  & 0.281 & 0.467 \\ \hline
 		 3 	& 0.664  & 0.779  & 0.514  & 0.667  & 0.467 & 0.628   & 0.572 & 0.669 		 & 0.484 & 0.628  & 0.579 & \textit{\textbf{0.719}}  & 0.284 & 0.477  \\ \hline
		4 	&0.679  & 0.771 & 0.543 & 0.714 & 0.544 & 0.606   &  0.607 & \textit{\textbf{0.772}} 	& 0.512 & 0.635 &  0.604 & \textit{\textbf{0.716}} & 0.309 & 0.474  \\ \hline
		5 	 &  0.579 &  0.729 & 0.438 & 0.686 & 0.467 & 0.656 &  0.566 & 0.724	& 0.453 & 0.639 & 0.544 & 0.691 & 0.312 &  0.484 \\ \hline
		Best Lag  	 &0.679 & 0.750  & 0.638 & \textit{\textbf{0.733}}  & 0.572 & \textit{\textbf{0.661}} & 0.634 & 0.731        & 0.495 & \textit{\textbf{0.660}} & 0.604 &0.705  & 0.365 & \textit{\textbf{0.512}} \\ \hline
		\tabincell{c}{All Lag (1-5)}   &0.643  &0.740  & 0.505 & 0.659 &0.506  & 0.639 & 0.628 & 0.717       & 0.474  & 0.596 & 0.649  & \textit{\textbf{0.716}} & 0.312 & 0.484  \\ \hline 
	\end{tabular}
	\end{adjustbox}

\end{table*}

\begin{table}[!ht]
	\centering
	\caption{Confusion matrix for multi-class classification problem (each row correspond to test scenario and each column denote predicted class) using SVM (top) and Conv. NN (bottom).}
	\begin{adjustbox}{width=0.9\linewidth}
		\begin{tabular}{c|c|c|c|c||c|c|c|c}
			\hline
			&\multicolumn{4}{c||}{Shape} &\multicolumn{4}{c}{Shape + Kinematics}\\ \hline
            & GC & GI & FC & FI  & GC & GI & FC & FI \\
			\hline
			GC &\textbf{0.16} &0.29 &0.18 &0.38 &\textbf{0.40} &0.27 &0.09 &0.24  \\
			GI &0.11 &\textbf{0.23} &0.26 &0.40 &0.03 &\textbf{0.77} &0.08 &0.12  \\
			FC&0.07 &0.38 &\textbf{0.32} &0.23  &0.02 &0.35 &\textbf{0.43} &0.20\\
			FI&0.08 &0.34 &0.14 &\textbf{0.44} &0.15 &0.42 &0.08 &\textbf{0.34}\\
			\hline
			\hline
			&\multicolumn{4}{c||}{Shape} &\multicolumn{4}{c}{Shape + Kinematics}\\ \hline
            & GC & GI & FC & FI  & GC & GI & FC & FI \\
			\hline
			GC &\textbf{0.88} &0.11 &0 &0   &\textbf{0.92} &0.07 &0 &0\\
			GI & 0.02 &\textbf{0.79} &0.16 &0.02  &0.01 &\textbf{0.81} &0.17 &0\\
			FC & 0 &0 &\textbf{0.90} &0.10  &0 &0.01 &\textbf{0.95} &0.03\\
			FI & 0.21 &0 &0.05 &\textbf{0.73} &0.14 &0.01 &0.02 &\textbf{0.82}\\
			\hline
		\end{tabular}
	\end{adjustbox}
	\label{tab:confisionmatrix2class}
\end{table}

\begin{table*}[h!]
	\centering
	\caption{The average classification rates for different experimental condition.} 
	\label{tab:class_real_two_class}
	 \begin{adjustbox}{width=\textwidth}
	\begin{tabular}{|c|c||c|c|c||c|c|c||c|c|c||c|c|c||}		
		\hline
		\multirow{3}{*}{Classifier}   & \multirow{3}{*}{\tabincell{c} {VAR \\ Model \\ Lag}} & \multicolumn{12}{c||}{Class. Rate (\%)}    \\ 
		\cline{3-14}  
		&  & \multicolumn{3}{c||}{\textbf{A.} GC vs. GI } & \multicolumn{3}{c||}{ \textbf{B.} GI vs FC } & \multicolumn{3}{c||}{ \textbf{C.} GI vs. FI } & \multicolumn{3}{c||}{ \textbf{D.} FC vs. FI} \\ 
		\cline{3-14} 
		&   &      Kine. & Shape & Comb. & Kine. & Shape & Comb. & Kine. & Shape & Comb.   & Kine. & Shape & Comb.     \\ 
		\hline
		\tabincell{c} {TSRVF-PCA \\+ Conv. NN} & -- & -- &0.623 & -- & -- & 0.568 & -- & -- & 0.533 & -- &--  & 0.598  & -- \\ \hline
		\tabincell{c} {TSRVF-PCA-VAR \\+ SVM} & -- & 0.686 & 0.679 & 0.793 & 0.610 & 0.638 & 0.733 & 0.550 & 0.572 & 0.661 &0.531  &0.634 & 0.772 \\ \hline
		\tabincell{c}{TSRVF-PCA-VAR \\+ Conv. NN} & Best Lag & \multirow{3}{*}{0.882} & 0.933 & 0.947 & \multirow{3}{*}{0.875} &0.900  & 0.844 & \multirow{3}{*}{0.896} &0.970  &0.970  &\multirow{3}{*}{0.877}  & 0.895  &0.923 \\ 
		\cline{1-2} \cline{4-5} \cline{7-8} \cline{10-11}  \cline{13-14} 
		\tabincell{c}{TSRVF-PCA-VAR \\+ Conv. NN} & \tabincell{c}{All Lag\\(1-5)} &  & 0.904 & \textbf{0.952} & & \textbf{0.912} & \textbf{0.912} &  & 0.974 & \textbf{0.988} &  & 0.933 & \textbf{0.955} \\ \hline
		
	\end{tabular}
	\end{adjustbox}
\end{table*}

\subsection{Biological Interpretation}
    From the experimental results, we conclude that the Euclidean time series modeling of shape space (TSRVF-PCA-VAR) is significantly more efficient than just TSRVF-PCA features in distinguishing cell migration patterns. They are also more informative and robust compared to the cell kinematics features. We hypothesis that: (1) In the heterogeneous amoeba population, the shape dynamics of the amoeba may be difficult to classify (Table.~\ref{tab:confisionmatrix2class} in the presence of inhibitor, as there may be a slight change in the actin-cytoskeleton localization and more remarkable change in its dynamics and/or contraction. This inhibition can lead to a significant difference in cell kinematics but not a considerable change in its morphology. (2) The mis-classification of ROCK inhibited amoeba  on glass/fibronectin with that on fibronectin/glass without inhibitor (17\%/14\%) could be due to the presence of cells that are either dying or static and adopt approximately a spherical shape which is almost constant over time. (3) The low percentage (1\%) of amoeba on glass with inhibitor, which was mis-classified as amoeba on glass without inhibitor, indicates a strong effect of the inhibitor on the mode of displacement of the amoeba.  
    (4) Similarly, the inhibitor affects the migration on fibronectin, as only 2\% of the amoeba are mis-classified as not inhibited amoeba on fibronectin. 

The results suggest that cytoskeleton dynamics of a \textit{E. histolytica} may be stimulated by different signaling pathways, depending on the support (glass and fibronectin). It also indicates that \textit{E. histolytica} is at least able to adopt migration modes that are different in the presence or absence of the ROCK inhibitor. Further biological experiments will reveal the effects of key biological players causing these differences.

\section{Conclusion}

The method developed in this paper employs the mathematical foundations of functional analysis to transport non-linear shapes to a Euclidean space. Subsequently, it fits a vector autoregressive (VAR) model to this time series to capture the dynamics of shape evolution. The strengths of this representation and modeling are demonstrated using tasks such as synthesis, prediction, and classification.  The designed shape features are sufficient to distinguish between amoeboid cell migrations on glass and fibronectin. Also, they are found to be more robust than kinematics features. With this framework, we can differentiate the migrating behavior of a cell by its shape evolution. The application of the method is not limited to shape classification but can also be utilized for shape synthesis and prediction, as demonstrated in this paper. To our knowledge, this is the first time where time-series modeling of dynamic shapes has been proposed and applied to solve a fundamental problem in biology. 

\appendices
	\section{DCC-GARCH Model}
	\label{app: garch_model}
	DCC-GARCH model is defined in Equation~\ref{eqn: garch}. Here $D(\tau)$ is a $d \times d$ diagonal matrix with diagonal entries $\sqrt{k_i(\tau)}$ and where $k_i(\tau) = w_{i,0} + \sum_{m=1}^{M_i}w_{i,m}l^2_i(\tau-m) + \sum_{n=1}^{N_i}\zeta_{i,k}k_i(\tau-n)$, for $i = 1,...,d$. $\Lambda(\tau)$ is a time-varying conditional-correlation matrix of the standardized residuals $\varepsilon(\tau)$.
	In DCC-GARCH model, the term $\Lambda(\tau)$ has additional structure that constrains its estimation from the data.  
	\begin{equation}
		\begin{split}
			\Lambda(\tau) &= Q_d{^-1}(\tau) Q(\tau) Q_d^{-1}(\tau) , \\
			Q(\tau) &= (1- a- b)\bar{Q}(\tau) + a \epsilon(\tau-1)\epsilon^{\dagger}(\tau-1) +bQ(\tau - 1),
		\end{split}
	\end{equation}
	where: 
	\begin{itemize}
		\item $\bar{Q}(\tau)$ is the unconditional covariance matrix of the standardized residuals $\varepsilon(\tau)$,
		\item $Q_d(\tau) = diag(\sqrt{q_{11}(\tau)},...,\sqrt{q_{dd}(\tau)})$ is a diagonal matrix with the square root of the diagonal elements of $Q(\tau)$.
		\item The parameter a and b are scalers.
	\end{itemize}
	We skip further details and refer the reader to \cite{Engle2002Dynamic}. 
	
	\section{Estimation of DCC-GARCH Model}
	\label{app: garch_estimation}
	The parameter set $\Theta$ consists of two subsets $(\boldsymbol{\phi}, \boldsymbol{\psi}) = (\phi_1,...,\phi_d,\psi)$, where $\phi_i = (w_{0,i},..., w_{M_i, i}, \zeta_{1,i}, ..., \zeta_{N_i, i})$ correspond to the parameters of the univariate GARCH model for the $i^{th}$ PCA component. In the first stage, the elements of $\boldsymbol{\phi}$ are estimated by maximizing quasi log-likelihood function: $QL_1(\boldsymbol{\phi}|l(\tau))$
	\begin{equation}
		= -\dfrac{1}{2} \sum_{i=1}^{d} \left( T \log(2\pi) + \sum_{\tau=1}^{T}\left( \log(k_i(\tau)) + \dfrac{l_i^2(\tau)}{k_i(\tau)} \right) \right).
	\end{equation}
	In the second stage the parameter $\boldsymbol{\psi}= (a, b)$ of dynamic correlation is estimated using the correctly specified log-likelihood. Dropping the terms that are constant in these parameters, we maximize: $	QL_2^*(\boldsymbol{\psi}|\hat{\boldsymbol{\phi}}, l(\tau))$
	\begin{equation}
		= -\dfrac{1}{2} \sum_{\tau=1}^{T} \left(\log(|\Lambda(\tau)|) + \epsilon^{\dagger}(\tau) \Lambda^{-1}(\tau) \epsilon_\tau \right).
	\end{equation}
	The details are presented in the textbook~\cite{Engle2002Dynamic}.
	
	\begin{table}[!htbp]
		\centering
		\caption{Estimation results for DCC-GARCH model for TSRVF-PCA representation of a shape sequence.} 
		\label{tab: garch_estimation}
		\begin{tabular}{|c|c|c|c|c|}
			\hline
			Parameter & Estimate & Std. Error & t value & $Pr(>|t|)$ \\
			\hline
			$w_{11}$   &   0.430  &  0.142 &  3.026 & 0.003 \\
			\hline
			$\eta_{11}$  &  0.993 &   0.002 & 524.454 & 0.000 \\
			\hline
			\multicolumn{5}{|c|}{$\cdots$} \\
			\hline
			$w_{51}$   &   0.520  &  0.110 &  4.738 & 0.000 \\
			\hline
			$\eta_{51}$   &  0.997   &  0.002 &  413.333  & 0.000 \\
			\hline
			$a$  &  0.044  &  0.016 &  3.037 & 0.002 \\
			\hline
			$b$ &  0.513 &  0.165 &  3.116 & 0.002 \\			
			\hline
		\end{tabular}			
	\end{table}
\vspace{-2pt}

\bibliographystyle{IEEEtran}	
\bibliography{MyRef_amoeba,bibfile}	
	
\end{document}